\documentclass[aps,prd,11pt,notitlepage]{revtex4-1}
\usepackage[english]{babel}
\usepackage{amsmath,amssymb,graphicx,bm}
\usepackage[colorlinks=true, linkcolor=red, citecolor=blue]{hyperref}
\usepackage[justification=raggedright,font=small,labelfont=bf]{caption}

\begin{document}

\title{Photon ring auto-correlations from gravitational fluctuations around a black hole}
\author{Qing-Hua Zhu}
\email{zhuqh@cqu.edu.cn}
\affiliation{Department of Physics, Chongqing University, Chongqing 401331, China}

\begin{abstract}
  The images of supermassive black holes in M87 and our galaxy captured by the Event Horizon Telescope (EHT) might open up a new way for exploring black hole physics at the horizon scale. Theoretically, this could provide insights into studying the emission sources around a black hole or the geometries of the black hole itself.
This paper investigates the two-point correlations of intensity fluctuations on the photon ring, resulting from the existence of shock waves or gravitational fluctuations around a black hole. Following approaches used in the field of gravitational wave detectors, we introduce response functions of Very Long Baseline Interferometry (VLBI) for detecting gravitational fluctuations and study the shape of the overlap reduction functions. It is found that the shape of the correlations here differs from that resulting from stochastic emission sources. 
By providing an order-of-magnitude estimate for the signal-to-noise ratio (SNR), we obtain sensitivity curves for the detection of gravitational fluctuations. This might reveal the potential for detecting gravitational fluctuations with future VLBI observations in the LIGO frequency band.

%  The images of supermassive black holes in M87 and our galaxy taken by Event Horizon Telescope (EHT) might open up a new window for studying black hole physics at horizon scale. Theoretically, it can provide a way for studying the emission sources around a black hole or the black hole geometries itself.  This paper investigates the two-point correlations of intensity fluctuations on the photon ring, as a result of existence of shock waves or gravitational fluctuations around a black hole. Following the approaches used in field of gravitational wave detectors, we introduce response functions of Very Long Baseline Interferometry (VLBI) for detecting the gravitational fluctuations, and study shape of the overlap reduction functions. It is found that the shape of the correlations here is different from that resulted from stochastic emission sources.  By providing an order-of-magnitude estimate for the signal-to-noise ratio (SNR), we obtain sensitivity curves for the gravitational fluctuation detection. It might reveal the potential for detecting the gravitational fluctuations with a future VLBI in the LIGO frequency band. 
\end{abstract}
\maketitle
\section{Introduction}

The images of the supermassive black holes in M87 and our galaxy taken by Event
Horizon Telescope (EHT) might open up a new window for studying black hole
physics at horizon scale, and potentially test gravity theory in the strong gravitation regime \cite{EventHorizonTelescope:2019dse,EventHorizonTelescope:2019ggy,EventHorizonTelescope:2019pgp,EventHorizonTelescope:2021dqv,EventHorizonTelescope:2022xnr}. Recently, a question was carefully addressed that how much physical information about the black hole physics can be extracted from the images. In principle, the images can reflect either i) black hole geometries \cite{Johannsen:2015hib,Vagnozzi:2022moj,Glampedakis:2021oie}, such as spins \cite{Bardeen:1973tla}, hairs \cite{Johannsen:2010ru,Cunha:2015yba,Gan:2021xdl,Ghosh:2022mka,Rosa:2022tfv}, and even cosmological constant \cite{Chowdhuri:2020ipb,Perlick:2021aok,Chang:2020miq}, or ii) different emission sources near the black hole, including Doppler effect of emission and different shapes of emission region \cite{Gralla:2019xty,Narayan:2019imo,Bronzwaer:2021lzo}. Theoretically, different emission models or space-time geometries have imprints on the images in a different way \cite{Kocherlakota:2022jnz}. The photon ring as a closed, convex brightness curve on the image plane is demonstrated to be insensitive the emission models \cite{Johannsen:2015hib,Narayan:2019imo,Gralla:2020srx,Gralla:2020yvo,Bronzwaer:2021lzo}. 
And the light rays from different lensing bands can also produce different order images of the emission sources, known as photon sub-rings
\cite{Johnson:2019ljv,Broderick:2021ohx,Wielgus:2021peu,Tsupko:2022kwi}. These observables are well-suited for investigating black hole geometries. However, limited by current resolution of EHT observations, these photon rings did not appear on the images  \cite{Gralla:2020pra,Vincent:2020dij,Tiede:2022grp}. One has to study black hole geometries with assumptions of emission models, and vice versa. 
Recently, a study on the appearance of photon rings in the presence of stochastic emission was also considered \cite{Hadar:2020fda}.
The two-point correlation of intensity fluctuation on the photon ring was introduced aiming for separating emission models from black hole background geometries.
 
In this paper, we investigate the two-point correlations of intensity fluctuations on the photon ring. Differed from the pioneers' study \cite{Hadar:2020fda,Chen:2022kzv}, %that ascribed the fluctuations to stochastic emission sources,
 we consider the intensity fluctuations caused by  gravitational fluctuations around a black hole. Since the observed intensity is given by $I_{\rm obs}=(1+z)^3 I_{\rm emt}$, both the emission sources and the trajectories of light can be stochastic. In this sense, the observed intensity contrast ${\delta I_{\rm obs}}/{I_{\rm obs}}$ can result from two contributions, namely,
\begin{eqnarray}
  \frac{\delta I_{\rm obs}}{I_{\rm obs}}=\frac{\delta I_{\rm emt}}{I_{\rm emt}}+3\left(\frac{\delta z}{1+z}\right)~.\label{0}
\end{eqnarray}
Since the $z$ is gravitational redshift, the gravitational fluctuations around a black hole could affect the propagation of light from emission regions to the locations of distant observers, and thus give rise to the redshift fluctuation $\delta z$.
If the two-point correlations of ${\delta I_{\rm obs}}/{I_{\rm obs}}$ in terms of ${\delta I_{\rm emt}}/{I_{\rm emt}}$ and $\delta z/(1+z)$ have different signatures, it can distinguish between the effects originating from emission sources and those arising from black hole geometries, in principle.

The second motivation of present paper is exploring the possibility of detecting stochastic gravitational fluctuations around a black hole with VLBI \cite{Sarkar:2021djs,Lin:2022ksb,Zhu:2022shb}. For example, the soft hairs can be implanted by sending a shock wave into a black hole \cite{Hawking:2016msc}. The shock wave as a type of propagating disturbance could exist in the vacuum permanently, which is analogous to the gravitational memory effects \cite{Strominger:2014pwa,Compere:2016hzt,Hawking:2016sgy}. Therefore, if the soft hair does exist, there are inevitably stochastic gravitational fluctuations consisting of the shock waves around a black hole. %In principle, it has influence on propagation of light, thereby potentially being detected with VLBI.
And quantifying the imprints left by the gravitational fluctuations on images would be non-trivial.

%It would be difficult
%We introduce for detecting gravitational fluctuations near a black hole. in which the  black hole and fluctuation are the perturbation propagation in the vacuum.
%It would be of interest for the latter one, because the gravitational fluctuations can reflect various physical systems, such as the extreme mass ratio binaries, primordial in the accrete disk.

%Since the photon rings are the assemble of
%the projection image of light rays from the photon sphere. The frequency fluctuations of two distinguished light rays from photon rings have possible to be correlated, because of space-time fluctuations have the same effect on the light rays. It seems not a new idea that investigates the shadow of black holes in the background of gravitational waves. The observable introduced is here frequency cross-correlations of light rays. By observing the lights, it also indirect observation on black hole perturbations, in additional to the flux of gravitational waves.
 
The rest of the paper is organized as follows. In Sec.~\ref{II}, we utilize the exact solutions of Regge-Wheeler equations \cite{Fiziev:2005ki} for describing gravitational fluctuations propagating in the vacuum. In order to study the shock waves, statistically, we introduce two-point correlations of the gravitational fluctuations. In Sec.~\ref{III}, we compute photon ring auto-correlations caused by the gravitational fluctuations, following the approaches for gravitational wave detectors \cite{Romano:2016dpx}, and study shape of overlap reduction functions. %It will be shown that the photon ring auto-correlations contributed by the redshift fluctuations is different from that ascribed to the stochastic emission. 
In Sec.~\ref{IVa}, we provide an order-of-magnitude estimate  for the observability of gravitational fluctuation detection with VLBI.
In Sec.~\ref{IV}, conclusions and discussions are summarized.

\section{Gravitational fluctuations in a Schwarzschild background} \label{II}

%Although, the dynamical evolution of shock waves produced by the soft hairs was explicitly given by diffeomorphism \cite{Hawking:2016sgy}, 
In order to describe all the types of shock waves freely propagating in a black hole background, we utilize the exact solutions of linear vacuum Einstein field equations in Schwarzschild background obtained by Fiziev \cite{Fiziev:2005ki}. In this context, the shock waves are no longer deterministic. To investigate these shock waves statistically,  we will introduce the two-point correlations of the gravitational fluctuations. 

In Einstein's theory of gravity, the Schwarzschild metric as a description of a spherical black hole is given by
\begin{eqnarray}
  {\rm d} s^2 & \equiv & \bar{g}_{\mu \nu} {\rm d} x^{\mu} {\rm d} x^{\nu}
  \nonumber\\
  & = & - A (r) {\rm d} t^2 + B (r) {\rm d} r^2 + r^2 {\rm d} \Omega^2
  \label{1}~,
\end{eqnarray}
where $A (r) = B^{- 1} (r) = 1 - {r_{\rm h}}/{r}$, and the $r_{\rm h}$ is
Schwarzschild radius. For quantifying gravitational fluctuation around the black hole, one can introduce the perturbed metric in the form of
\begin{eqnarray}
  g_{\mu \nu} & = & \bar{g}_{\mu \nu} + \delta g_{\mu \nu}  ~, \label{1a}
\end{eqnarray}
where the metric perturbation $\delta g_{\mu \nu}$ in
Regge-Wheeler gauge is \cite{Maggiore:2018sht}
\begin{eqnarray}
  \delta g_{\mu \nu}^{\rm{RW}} & = & \sum_{l = 2}^{\infty} \sum_{m = - l}^l
  \left( - \frac{1}{r} \sqrt{2 l (l + 1)} h^{B  t} \bm{T}_{l
    m, \mu \nu}^{B  t} - \frac{1}{r} \sqrt{2 l (l + 1)} h^{B
    1} \bm{T}_{l  m, \mu \nu}^{B  1} \right)
  \nonumber\\
  & = & \sum_{l = 2}^{\infty} {\sum_{m = - l}^l}  \left(\begin{array}{cccc}
      0    & 0    & h^{B  t}_{l  m} \csc \theta \partial_{\phi} & -
      h^{B  t}_{l  m} \sin \theta \partial_{\theta}                      \\
      0    & 0    & h^{B 1}_{l  m} \csc \theta \partial_{\phi}  & - h^{B
      1}_{l  m} \sin \theta \partial_{\theta}                            \\
      \ast & \ast & 0                                           & 0      \\
      \ast & \ast & 0                                           & 0
    \end{array}\right) Y_{l  m} ~ .  \label{2}
\end{eqnarray}
The $Y_{l  m}$ is spherical harmonics, and $\bm{T}_{l  m, \mu
    \nu}^{B  t}$ and $\bm{T}_{l  m, \mu \nu}^{B  1}$
are  Zerilli tensor spherical harmonics. In principle, the expansion of above metric perturbations is defined with $l\geq1$. We here consider the cases of $l \geqslant 2$ in Eq.~(\ref{2}), because only gravitational waves can carry the energy propagating to the distant observers.
 %We would explain latter for the $l\geq2$ that we adopted in Eq.~(\ref{2}).

Based on Einstein field equations, the equations governing the evolution of the metric perturbations  are given by
\begin{eqnarray}
  0  =  - \bar{\nabla}_b \bar{\nabla}_a \delta g^c_{\hspace{.5em} c} +
  \bar{\nabla}_c \bar{\nabla}_a \delta g_b^{\hspace{.5em} c} + \bar{\nabla}_c
  \bar{\nabla}_b \delta g_a^{\hspace{.5em} c} - \bar{\nabla}_c \bar{\nabla}^c
  \delta g_{a  b} + \bar{g}_{a  b} \left( \bar{\nabla}_d
  \bar{\nabla}^d \delta g^c_{\hspace{.5em} c} - \bar{\nabla}_d \bar{\nabla}_c
  \delta g^{c  d} \right) ~,
\end{eqnarray}
where the $\bar{\nabla}_{\mu}$ is the covariant derivative with respect to the
Schwarzschild background. Substituting the $\delta g_{\mu \nu}$ with the
metric perturbations in Eq.~(\ref{2}) and separating variables, one might
find $h^{B 1}_{l  m}$, $h^{B  t}_{l  m}$ can be
expressed in terms of one quantity $Q _{l  m}$, namely,
\begin{eqnarray}
  h^{B 1}_{l  m} & = & r \sqrt{\frac{B}{A}} Q_{l  m} ~,
  \label{4}\\
  \partial_t h^{B  t}_{l  m} & = & \sqrt{\frac{A}{B}}
  \partial_r (r  Q_{l  m}) ~,  \label{5}
\end{eqnarray}
and the $Q_{l  m}$ satisfies the Regge-Wheeler equation,
\begin{eqnarray}
  \partial_t^2 Q_{l  m} - \left( 1 - \frac{r_{\rm h}}{r} \right) \partial_r
  \left( \left( 1 - \frac{r_{\rm h}}{r} \right) \partial_r Q_{l  m} \right)
  + A \left( \frac{l (l + 1)}{r^2} - \frac{3 r_{\rm h}}{r^3} \right) Q_{l
      m} & = & 0 ~.  \label{6}
\end{eqnarray}
Based on Ref.~\cite{Fiziev:2005ki}, we adopt the exact solutions of Regge-Wheeler
equations in the form of
\begin{eqnarray}
  Q_{l  m} & = & \int \frac{{\rm d} \omega}{2 \pi}  \mathcal{A}_{l
    m} (\omega) e^{- i  \omega (t - r)} \left( \frac{r}{r_{\rm h}} - 1
  \right)^{i  r_{\rm h} \omega} \left( \frac{r}{r_{\infty}} \right)^3 \nonumber\\
  && \times
  \frac{{\rm{HeunC}} \left[ l (l + 1) - 6 - 6 i  r_{\rm h} \omega, - 6 i
  r_{\rm h} \omega, 1 + 2 i  r_{\rm h} \omega, 5, - 2 i  r_{\rm h} \omega, 1 -  {r}/{r_{\rm h}}
  \right]}{{\rm{HeunC}} [l (l + 1) - 6 - 6 i  r_{\rm h} \omega, - 6 i
  r_{\rm h} \omega, 1 + 2 i  r_{\rm h} \omega, 5, - 2 i  r_{\rm h} \omega, 1 - r_{\infty}/r_{\rm h}]}
  ~,  \label{7}
\end{eqnarray}
where the $\rm{HeunC}$ denotes confluent Heun's function, and the $\omega$ is frequency defined via $Q_{lm}(t,r)=(2\pi)^{-1}\int {\rm d}\omega Q_{lm,\omega}(r)e^{-i \omega t}$. In the asymptotic flatness regime, the $Q_{l  m}$ reduces to
\begin{eqnarray}
  Q_{l  m} (t, r \rightarrow \infty) & = & \int \frac{{\rm d} \omega}{2 \pi}
  \mathcal{A}_{l  m} (\omega) e^{- i  \omega (t - r)} ~. \label{8}
\end{eqnarray}
It is the up-mode solution of wave equations. The $\mathcal{A}_{lm}(\omega)$ encodes physical information of gravitational fluctuations at spatial infinity.
\begin{figure}
  \includegraphics[width=1\linewidth]{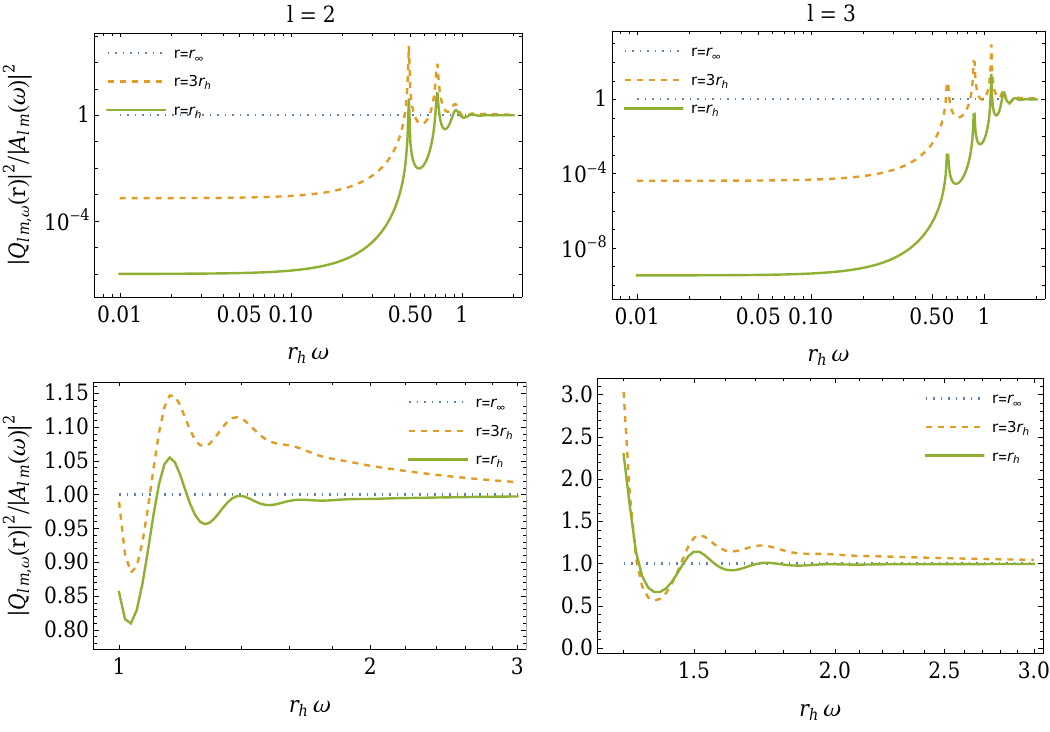}
  \caption{The $|Q_{lm,\omega}(r)|^2/|\mathcal{A}_{lm}(\omega)|^2$ as function of frequency $r_{\rm h}\omega$ for selected $r$ and multipole $l$. Top panels show the range of frequency from $0.01$ to $2$, and bottom panels show the range of the frequency larger than $1$. %Left panels and right panels shows the multipole $l=2$ and $l=3$, respectively. 
  Here, we have let the $r_{\infty}=10r_{\rm h}$ for illustrations.
  \label{F0a}}
\end{figure}
In Fig.~\ref{F0a}, we show the $Q_{lm, \omega}(r) $ as function of frequency $\omega$ for given radius and multipole. The amplitude of gravitational fluctuations are suppressed in the low-frequency regime.
To obtain the physical meaning of $\mathcal{A}_{l  m} (\omega)$, one should
transform the metric perturbation $\delta g^{\rm{RW}}_{\mu \nu}$ into the transverse-traceless (TT)
gauge, namely
\begin{eqnarray}
  \delta g^{\rm{RW}}_{\mu \nu} & = & \delta g^{{\rm T T}}_{\mu \nu}
  +\mathcal{L}_{\xi} \bar{g}_{\mu \nu} ~,  \label{9}
\end{eqnarray}
Eliminating the $h^{B  t, {\rm T T}}_{l  m}$ in
Eq.~(\ref{9}) with given infinitesimal vector $\xi^{\mu}$ \cite{Maggiore:2018sht}, one can
obtain
\begin{eqnarray}
  h^{B  t}_{l  m} & = &  \partial_t h^{B 2, {\rm T T}}_{l
      m} ~,  \label{10}\\
  h^{B 1}_{lm} & = & h^{B 1, {\rm T T}}_{l  m} + \left( \partial_r
  - \frac{2}{r} \right) h^{B 2, {\rm T T}}_{l  m} ~,
  \label{11}
\end{eqnarray}
where $h^{B 2,\rm TT}_{l  m}$ is the spin-2 component of metric
perturbation. Using Eqs.~(\ref{9})--(\ref{11}), the metric
perturbations in TT gauge are evaluated to be
\begin{eqnarray}
  \delta g_{\mu \nu}^{{\rm T T}} & = & \sum_{l = 2}^{\infty} \sum_{m = - l}^l
  \left( - \frac{1}{r} \sqrt{2 l (l + 1)} h^{B  1,\rm TT}_{l
    m} \bm{T}_{l  m, \mu \nu}^{B  1} + \frac{1}{r^2}
  \sqrt{\frac{2 (l + 2) !}{(l - 2) !}} h^{B 2,\rm TT}_{l  m}
  \bm{T}^{B 2}_{l  m, \mu \nu} \right) ~, \label{14a}
\end{eqnarray}
where $\bm{T}^{B 2}_{l  m, \mu \nu}$ is Zerilli tensor harmonics
with spin-2. Based on Eqs.~(\ref{4}), (\ref{5}) and (\ref{14a}), the $\delta g_{\mu \nu}^{{\rm T T}}$ in the asymptotic flatness regime reduces to
\begin{eqnarray}
  \delta g_{\mu \nu}^{{\rm T T}} |_{r \rightarrow \infty} & =  & \sum_{l =
    2}^{\infty} \sum_{m = - l}^l \int \frac{{\rm d} \omega}{2 \pi} \left\{
  \frac{1}{i  \omega  r} \sqrt{\frac{2 (l + 2) !}{(l - 2) !}}
  \mathcal{A}_{l  m} (\omega) e^{- i  \omega  (t - r)} \right\}
  \bm{T}^{B 2}_{l  m, \mu \nu} +\mathcal{O} \left( \frac{1}{r^2}
  \right) ~ . \label{13}
\end{eqnarray}
The leading order of $\delta g_{\mu \nu}^{{\rm T T}}$ at $r\rightarrow \infty$ is also known as gravitational waves, because it can carry energy into spatial infinity. Using Eq.~(\ref{13}), the radiated power for the gravitational waves is given by
\begin{eqnarray}
  P & = & \frac{1}{32 \pi} \int r^2 {\rm d} \Omega \langle \dot{\delta g}_{i
    j}^{{\rm T T}} \dot{\delta g}_{i  j}^{{\rm T T}} \rangle
  \nonumber\\
  & = & \frac{1}{32 \pi} \int \frac{{\rm d} \omega}{2 \pi} \int \frac{{\rm d} \omega'}{2
    \pi} \sum_{l, m, l', m'} \sqrt{\frac{2 (l + 2) !}{(l - 2) !} \frac{2 (l' +
      2) !}{(l' - 2) !}} e^{- i (\omega - \omega') (t - r) } \langle \mathcal{A}_{l
    m} (\omega) \mathcal{A}_{l'  m'}^{*} (\omega') \rangle \int
  {\rm d} \Omega \bm{T}^{B 2}_{l  m, i  j} \bm{T}^{B
    2}_{l'  m', i  j} \nonumber\\
  & = & \frac{1}{16 \pi} \int \frac{{\rm d} \omega}{2 \pi} \sum_{l = 2}^{\infty}
  \sum_{m = - l}^l \frac{(l + 2) !}{(l - 2) !} P_{\mathcal{A},lm}(\omega) ~,  \label{14}
\end{eqnarray}
where we have used orthogonal condition $\int
  {\rm d} \Omega \bm{T}^{B 2}_{l  m, i  j} \bm{T}^{B
    2}_{l'  m', i  j} = \delta_{l  l'} \delta_{m
    m'}$, and the temporal average $\langle \mathcal{A}_{l  m} (\omega) \mathcal{A}_{l'
    m'}^{*} (\omega') \rangle = 2\pi \delta(\omega-\omega')\langle \mathcal{A}_{l  m} (\omega) \mathcal{A}_{l'
    m'}^{*} (\omega) \rangle$. Because the radiated power is determined by the $\mathcal{A}_{lm}$ with $l = l'$ and $m = m'$, we thus give two-point correlations of the $\mathcal{A}_{lm}(\omega)$ %for describing stochastic gravitational fluctuations at spatial infinity  
in the form of 
\begin{eqnarray}
  \langle \mathcal{A}_{l  m} (\omega) \mathcal{A}_{l'
    m'}^{*} (\omega') \rangle & = & 2 \pi \delta_{l  l'} \delta_{m
    m'} \delta (\omega - \omega') P_{\mathcal{A},lm}(\omega) ~.
  \label{15}
\end{eqnarray}
%we adopt the angular average with $\delta_{l  l'} \delta_{m m'}$ in Eq.~(\ref{15}). In principle, one gives a general form of $\langle \mathcal{A}_{l  m} (\omega) \mathcal{A}_{l' m'}^{*} (\omega') \rangle = 2 \pi C_{l  l' m' m'} \delta (\omega - \omega') |\mathcal{A}_{l  m} (\omega) |^2$ and $C_{l  l, m  m} = 1$.In this case, the explicity expression of $C_{l  l' m' m'}$ can not be known from observing the radiated power of the gravitational waves.
%With above assumptions, it also indicates that one can not extract polarization of  gravitational waves by using the two-point correlations. 
%It is consistent with the argument that 
%It is based on the assumptions that the radiated power of gravitational perturbations are unresolvable and encoded in the stochastic variable $\mathcal{A}_{ lm}(\omega)$. 
Here, the $\mathcal{A}_{lm}(\omega)$ is assumed to be an stochastic variable for describing shock waves in the Schwarzschild background.
Through introducing radiated power spectrum $P_{l  m} (\omega)$ with
$P =  \int \frac{{\rm d} \omega}{2 \pi} \sum_{l, m} P_{l  m} (\omega)$,
we have
  $P_{\mathcal{A},lm}(\omega)  =  16 \pi {(l - 2) !}({(l + 2)
    !})^{-1} P_{l  m} (\omega) ~. $ 
Therefore, based on Eq.~(\ref{15}), one can introduce
\begin{eqnarray}
  h^{\rm{GW}}_{l  m} (\omega) & \equiv & \frac{1}{4} \sqrt{\frac{(l + 2)
      !}{\pi (l - 2) !}} \mathcal{A}_{l  m} (\omega) ~ ,  \label{18}
\end{eqnarray}
such that two-point correlations of $h^{\rm{GW}}_{l  m} (\omega)$ can be expressed in terms of the radiated power, namely,
\begin{eqnarray}
  \langle h_{l  m}^{\rm{GW}} (\omega) h_{l'  m'}^{\rm{GW}} (\omega')
  \rangle & = & 2 \pi \delta_{l  l'} \delta_{m  m'} \delta (\omega
  - \omega') P_{l  m} (\omega) ~.  \label{19}
\end{eqnarray}
%The gravitational waves $h_{l  m}^{\rm{GW}} (\omega)$ with different $l, m$ and frequency $\omega$ are un-correlated. 
%Because the radiated power of gravitational waves could be a practical observable, we considered the metric perturbations with $l\geq2$ in Eq.~(\ref{2}).
It is expected that physical information of the shock waves is encoded in the stochastic variables $\mathcal{A}_{lm}(\omega)$ (or $h_{l  m}^{\rm{GW}} (\omega)$), as well as the transfer functions shown in Eq.~(\ref{7}). In principle, the shape of radiated power spectrum  $P_{l  m} (\omega)$ could be obtained from the gravitational wave detections. As the production mechanism of the shock waves is beyond the scope of the present paper, explicit expressions for $P_{l  m} (\omega)$ can not be provided here.
%We here can not provide the explicit expressions of $P_{l  m} (\omega)$, because production mechanism of the shock waves is not involved in the present paper.   
In spite of this, one can still extract physical information of the shock waves from the transfer functions. It will be addressed in the following sections.
%of detecting gravitational fluctuations near the black hole with Event Horizon Telescope

\section{Imprints of gravitational fluctuations on black hole images} \label{III}

%The images of black holes taken by EHT suggest that black hole geometries could be inferred from the images. 
Inspired by pioneers' studies \cite{Hadar:2020fda}, we consider observed intensity fluctuations on the photon ring, which is originated from gravitational fluctuations around a supermassive black hole. Based on $\delta I_{\rm obs}/I_{\rm obs} \propto 3 \delta z/(1+z)$ in Eq.~(\ref{0}), in this section, we  would relate the redshift fluctuations with the stochastic gravitational fluctuations, $h_{lm}^{\rm GW}$. %Since the fluctuations are assumed to be stochastic, the fluctuations are studied statistically with the two-point correlations. The approaches has been widely used in gravitational wave detectors in various frequency band for detecting stochastic background \cite{Romano:2016dpx}. Therefore, we in fact introduce an approach for detecting gravitational fluctuation near the black hole with the Event Horizon Telescope from the intensity images.

\begin{figure}
  \includegraphics[width=1\linewidth]{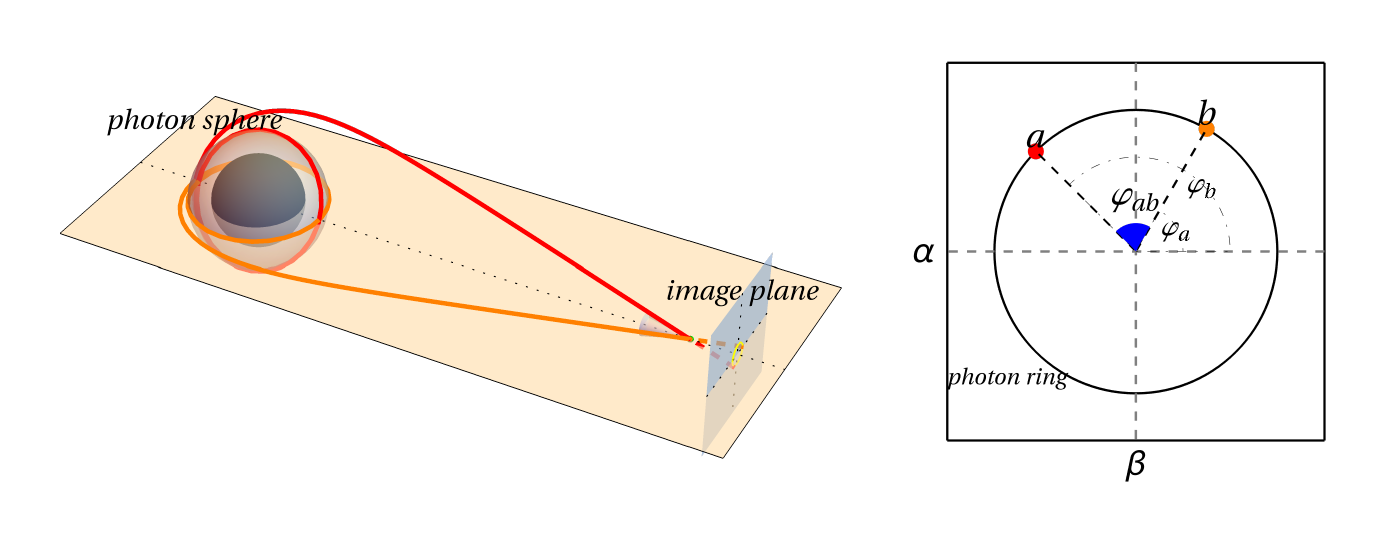}
  \caption{Left panel: schematic diagram of the light rays $a$ and $b$ propagating from photon sphere to observers' image plane. Right panel: schematic diagram of the light rays $a$ and $b$ on image plane. \label{F1}}
\end{figure}

Since the gravitational fluctuations are assumed to be stochastic for describing the shock waves, we follow the approaches that have been widely used in the field of gravitational wave detectors for detecting stochastic gravitational wave background \cite{Romano:2016dpx}. The schematic diagram is shown in Fig.~\ref{F1}.
The light rays $a$ and $b$ from photon sphere are projected onto image plane with angle $\varphi_{ab}$. In the present of gravitational fluctuations, the intensity fluctuations of the light rays $a$ and $b$ should be correlated. It is analogous to the studies of pulsar timing arrays as gravitational wave detectors \cite{Romano:2016dpx}, in which the two-point correlations of timing residuals of a pulsar pair is non-trivial, and can provide characteristic signatures known as Hellings-Downs curve \cite{Hellings:1983fr}.

%Recent observations on the supermassive black hole in our galaxy have release time series images of the black holes. In principle, the space-time fluctuations near the  lack holes also can be revealed by the optical observations in the future.

\subsection{Redshift fluctuations}

%The reshift fluctuations of light in the perturbed Schwarzschild black hole given in Sec.~\ref{II}.

The redshift of a photon from emission region to the detectors is defined with
\begin{eqnarray}
  1+z & = & \frac{(u_{\mu} p^{\mu})_{\rm{obs}}}{(u_{\mu}
    p^{\mu})_{\rm{emt}}} ~,  \label{20}
\end{eqnarray}
where the $u_{\mu}$ is 4-velocity of a reference observer, the $p^{\mu}$ is 4-momentum of
light, and the subscripts `obs' and `emt' denote the events of detectors and emission sources, respectively.
%However, in practical, the redshift might not be a good observable, because it might be difficult to known frequency of lights near the black hole, which could be emitted from the source even far from the black hole. There would be different situations for the frequency fluctuations, since the initial (emitted) frequency is not important for the relative variation of frequency.
Employing Eq.~(\ref{20}) and expanding the 4-velocity $u^{\mu}$ and 4-momentum 
$p_{\mu}$ in the form of
\begin{eqnarray}
  p^{\mu}  =  \bar{p}^{\mu} + \delta p^{\mu} ~, \text{and}~ u_{\nu}  =  \bar{u}_{\nu} + \delta u_{\nu} ~,  \label{21}
\end{eqnarray}
we obtain redshift fluctuations, namely
\begin{eqnarray}
  \frac{\delta z}{1+z} & = &  \left. \frac{\delta {u}_{\mu} p^{\mu} + \bar{u}_{\mu} \delta
    p^{\mu}}{\bar{u}_{\nu} p^{\nu}} \right|_{\rm{obs}} - \left. \frac{\delta u_{\mu}
    p^{\mu} + \bar{u}_{\mu} \delta p^{\mu}}{\bar{u}_{\nu} p^{\nu}} \right|_{\rm{emt}}
  ~ . \label{22}
\end{eqnarray}
%Because photon emitters near a black hole might move much slower than the speed of light, 
Here, we let $\bar{u}^\mu=\bar{u}^0\delta_0^\mu$ for simplicity.
In this case, the $\bar{u}^{\mu}$ is the normal vector
adapted to synchronous hypersurface for the coordinate time $t$. And thus one can obtain
the expansion of $u^{\mu}$, namely,
\begin{eqnarray}
  \bar{u}_{\nu} & = & - \sqrt{A} {\rm d} t ~, \\
  \delta u_{\nu} & = & 0 ~ .
\end{eqnarray}
The $\delta u_{\nu}  =  0$ results from $\delta g^{00}=0$ for the axial metric  perturbations. It shows that the perturbed 4-velocity $\delta u_\mu$ takes the same  form as those in RW gauge and TT gauge. Substituting the expression of 4-velocities into Eq.~(\ref{22}), we obtain the redshift fluctuations in the form of
\begin{eqnarray}
  \frac{\delta z}{1+z} & = & \left. \frac{\delta p^0}{p^0} \right|_{\rm{obs}}
  - \left. \frac{\delta p^0}{p^0} \right|_{\rm{emt}} ~ .  \label{27}
\end{eqnarray}
The $\delta z / (1+z)$ is determined by the difference of the $\delta p^0/p^0$, and thus does not explicitly depend on the metric. %One canonly depends on the time-components of the 4-momentum of lights. 

Here, we will show the imprints of gravitational fluctuations on the
$\delta z / (1+z)$.
Firstly, one should obtain how the light rays propagate in the Schwarzschild
background. It is based on geodesic equations, namely,
\begin{eqnarray}
  \bar{p}^{\mu} \bar{\nabla}_{\mu} \bar{p}^{\nu} & = & 0 ~ .
\end{eqnarray}
By making use of Hamilton-Jacobi method for solving the geodesic equations, the
solutions can be obtained in the form of
\begin{eqnarray}
  \bar{p}^0 & = & \frac{E}{A} ~,  \label{28}\\
  \bar{p}^r & = & E \sqrt{\frac{1}{B} \left( \frac{1}{A } -
    \frac{\kappa}{r^2} \right)} ~, \\
  \bar{p}^{\theta} & = & \frac{E}{r^2} \sqrt{\kappa - \frac{\lambda^2}{\sin^2
      \theta}} ~, \\
  \bar{p}^{\phi} & = & \frac{E \lambda}{r^2 \sin^2 \theta} ~,  \label{31}
\end{eqnarray}
where $E$, $\lambda$ and $\kappa$ are the constants of integration from the
geodesic equations. It can represent the energy of light at spatial
infinity, angular momentum per energy along $z$-axis, and square of total angular momentum per energy, respectively. 

The appearance of a black hole is closely related to property of the
photon sphere, which leads to a sharp brightness ring on the image planes, known as photon ring. The location of photon sphere $r_{\rm ps}$ is formulated by unstable photon orbits ${\rm d} p^r / {\rm d} r = p^r = 0$ that can be evaluated to be
$r_{\rm{ps}}  A' (r_{\rm{ps}}) - 2 A (r_{\rm{ps}})  =  0$.
%Namely, the $r_{\rm{ps}}$ can be obtained by solving above algebra equations. 
On the photon sphere, the $\lambda$ and $\kappa$ can be partly
determined, namely,
\begin{eqnarray}
  \kappa & = & \kappa_{\rm ps} \equiv \frac{r_{\rm{ps}}^2}{A (r_{\rm{ps}})} ~,  \label{34}\\
  \lambda & \leqslant & \sqrt{\kappa_{\rm ps}} \sin \theta_o ~,
\end{eqnarray}
where $\theta_o$ is the inclination angular of observers/detectors. The light rays reaching a distant observer/detector on the image plane can be described with the Cartesian coordinates $(\alpha, \beta)$, namely, \cite{Bardeen:1973tla}
\begin{eqnarray}
  \alpha & \equiv & - r_o^2 \sin \theta_o \left. \frac{p^{\phi}}{p^r}
  \right|_{(r_o, \theta_o)} = - \frac{\lambda}{\sin \theta_o} ~, \\
  \beta & \equiv & \pm r_o^2 \left. \frac{p^{\theta}}{p^r} \right|_{(r_o,
  \theta_o)} = \pm \sqrt{\kappa - \lambda^2 \csc^2 \theta_o} ~ .
\end{eqnarray}
One can also introduce polar angle$\begin{array}{l} \varphi \equiv \arcsin \left( {\alpha}/{\sqrt{\alpha^2 + \beta^2}}
    \right)
  \end{array}$ shown in schematic diagram in Fig.~\ref{F1}. And the $\lambda$ can be rewritten in terms of the polar angle, namely,
\begin{eqnarray}
  \lambda & = & - \sqrt{\kappa} \sin \theta_o \sin \varphi ~ .  \label{48}
\end{eqnarray}
The parameter pair $(\sqrt{\kappa}, \varphi)$ is the polar coordinate locating the observed light rays on image plane.

To study the redshift fluctuations of light based on Eq.~(\ref{27}), we utilize perturbed geodesic equations in the form of
\begin{eqnarray}
  \delta p^{\mu} \bar{\nabla}_{\mu} \bar{p}^{\nu} + \bar{p}^{\mu}
  \bar{\nabla}_{\mu} \delta p^{\nu} + g^{\nu \rho} \left( \bar{\nabla}_{\mu}
  \delta g_{\lambda \rho} - \frac{1}{2} \bar{\nabla}_{\rho} \delta g_{\mu
    \lambda} \right) \bar{p}^{\mu} \bar{p}^{\lambda} & = & 0 ~ .
\end{eqnarray}
It is derived from the expansions of geodesic equations $p^{\mu} \nabla_{\mu} p^{\nu} = 0 $ with Eqs.~(\ref{1a}) and (\ref{21}).
Using Eqs.~(\ref{1}) and (\ref{28})--(\ref{31}),  time component of the perturbed geodesic equations can be rewritten in the
form of
\begin{eqnarray}
  \bar{p}^{\mu} \partial_{\mu} \left( \frac{\delta p^0}{\bar{p}^0} \right) & =
  & - \frac{g^{00}}{\bar{p}^0} \left( \bar{\nabla}_{\mu} \delta g_{\lambda 0}
  - \frac{1}{2} \bar{\nabla}_0 \delta g_{\mu \lambda} \right) \bar{p}^{\mu}
  \bar{p}^{\lambda} ~.
\end{eqnarray}
It is found that the $\delta p^0$ is decoupled from spatial components of the 4-momentum $\delta p^{\mu}$. And above equation also can be given explicitly, namely,
\begin{eqnarray}
  &  & \left( \frac{1}{A} \partial_0 + \sqrt{\frac{1}{B} \left( \frac{1}{A} -
    \frac{\kappa}{r^2} \right)} \partial_r + \frac{ \sqrt{\kappa -
      \lambda^2 \csc^2 \theta}}{r^2} \partial_{\theta} + \frac{\lambda \csc^2
    \theta}{r^2} \partial_{\phi} \right) \left( \frac{\delta p^0}{p^0} \right)
  \nonumber\\
  %  & = & \sum_{l = 2}^{\infty} \sum_{m = - l}^l \left( \frac{\csc^2 \theta
  %  \left( \sqrt{\kappa - \lambda^2 \csc^2 \theta} \partial_{\phi} Y_{l 
  %  m} - \lambda \partial_{\theta} Y_{l  m} \right)}{r^3}
  %  \sqrt{\frac{r^2 - \kappa A}{A  B}} (\partial_r h^{B  t}_{l
  %   m} - \partial_t h^{B  1}_{l  m}) \right.\nonumber\\
  %  &&  \left. + \frac{\csc
  %  \theta}{r^4} \left( - \left( \cot \theta (\kappa - 2 \lambda^2 \csc^2
  %  \theta) + 2 \sqrt{\kappa - \lambda^2 \csc^2 \theta} \sqrt{\frac{r^2 - \kappa
  %  A}{A  B}} \right) \partial_{\phi} Y_{l  m} \right.\right.\nonumber\\
  %  &&  \left. \left.+ \lambda \csc^2
  %  \theta \sqrt{\kappa - \lambda^2 \csc^2 \theta} \partial_{\phi}^2 Y_{l
  %   m} + \lambda \left( \cot \theta \sqrt{\kappa - \lambda^2 \csc^2
  %  \theta} + 2 \sqrt{\frac{r^2 - \kappa A}{A  B}} \right)
  %  \partial_{\theta} Y_{l  m} \right.\right.\nonumber\\
  %  &&  \left. \left.+ (\kappa - 2 \lambda^2 \csc^2 \theta)
  %  \partial_{\theta} \partial_{\phi} Y_{l  m} - \lambda \sqrt{\kappa -
  %  \lambda^2 \csc^2 \theta} \partial_{\theta}^2 Y_{l  m} \right) h^{B
  %   t}_{l  m} \right) ~ . \label{42} \\
  & = & - \frac{\csc \theta}{r^4} \sum_{l = 2}^{\infty} \sum_{m = - l}^l \Bigg( \left( \sqrt{\kappa - \lambda^2 \csc^2
    \theta} \partial_{\phi} Y_{l  m} - \lambda \partial_{\theta} Y_{l
    m} \right) \sqrt{\frac{1}{B} \left( \frac{1}{A } -
    \frac{\kappa}{r^2} \right)} \partial_t h_{lm}^{B  1, {\rm TT}} \nonumber\\
  &&   +  \bigg( (\kappa - 2 \lambda^2 \csc^2 \theta) ( \partial_{\theta} \partial_{\phi} Y_{l  m} - \cot \theta \partial_{\phi}  Y_{l  m})  \nonumber\\
  &&   +
  \lambda \sqrt{\kappa - \lambda^2 \csc^2 \theta} (\csc^2 \theta
  \partial_{\phi}^2 Y_{l  m} + \cot \theta \partial_{\theta} Y_{l
    m} - \partial_{\theta}^2 Y_{l  m}) \bigg) \partial_t h_{lm}^{B
  2, {\rm TT}} \Bigg)~. \label{42}
\end{eqnarray}
%Here, we have adopted the metric perturbation $\delta g_{\mu \nu}$ in TT gauge. 
The first order differential partial equation can be solved with its characteristic equations, namely,
\begin{subequations}
\begin{eqnarray}
  \frac{{\rm d} \tau}{{\rm d} s} & = & \frac{1}{A}  \label{39}\\
  \frac{{\rm d} \rho}{{\rm d} s} & = & \sqrt{\frac{1}{B} \left( \frac{1}{A} -
    \frac{\bar{\kappa}}{\rho^2} \right)} \\
  \frac{{\rm d} \theta}{{\rm d} s} & = & \frac{ \sqrt{\bar{\kappa} -
      \bar{\lambda}^2 \csc^2 \theta}}{\rho^2} \\
  \frac{{\rm d} \phi}{{\rm d} s} & = & \frac{\bar{\lambda} \csc^2
    \theta}{\rho^2}  \\
  \frac{{\rm d}}{{\rm d} s} \left( \frac{\delta p^0}{p^0} \right) & = & \sum_{l
    = 2}^{\infty} \sum_{m = - l}^l \int \frac{{\rm d} \omega}{2 \pi} \mathcal{A}_{l
    m} ({\omega}) \mathcal{H}_{l  m,\omega} (\tau (s), \rho (s), \theta
  (s), \phi (s) ; \bar{\kappa}, \bar{\lambda})e^{-i \omega r_h \tau}  \label{43}
\end{eqnarray}\label{43a}
\end{subequations}
where we have used dimensionless variables $\tau \equiv t / r_{\rm h}$, $\rho \equiv
  r / r_{\rm h}$, $\bar{\kappa} \equiv \kappa / r_{\rm h}^2$, $\bar{\lambda} \equiv \lambda
  / r_{\rm h}$ in these equations. And by substituting Eqs.~(\ref{4}), (\ref{5}), (\ref{8}), (\ref{10}) and (\ref{11}) into right hand side of Eq.~(\ref{42}), we obtain the expression of $\mathcal{H}_{l  m, \omega}$ in Eq.~(\ref{43}), namely,
\begin{eqnarray}
  \mathcal{H}_{l  m, \omega} (\tau, \rho, \theta, \phi; \bar{\kappa}, \bar{\lambda}) & \equiv &  \frac{i   \csc \theta}{\omega  r_h        \rho^4}
  \Bigg(\left( \frac{l (l + 1) - 2}{\rho} \right) \sqrt{\frac{1}{B}
    \left( \frac{1}{A } - \frac{\bar{\kappa}}{\rho^2} \right)} \left(
  \sqrt{\bar{\kappa} - \bar{\lambda}^2 \csc^2 \theta} \partial_{\phi} Y_{l
    m} - \bar{\lambda} \partial_{\theta} Y_{l  m} \right) \left( \frac{ Q_{l  m , \omega}}{\mathcal{A}_{lm}(\omega)}\right) \nonumber\\
  &  & - \left( 1 - \frac{1}{\rho} \right)  \Big((\bar{\kappa} - 2
  \bar{\lambda}^2 \csc^2 \theta) (- \cot \theta \partial_{\phi} Y_{l
      m} + \partial_{\theta} \partial_{\phi} Y_{l  m})\nonumber\\
  &  & + \left. \bar{\lambda} \sqrt{\bar{\kappa} - \bar{\lambda}^2 \csc^2
    \theta} (\csc^2 \theta \partial_{\phi}^2 Y_{l  m} + \cot \theta
  \partial_{\theta} Y_{l  m} - \partial_{\theta}^2 Y_{l  m})
  \right) \partial_{\rho} \left( \frac{\rho Q_{l  m , \omega}}{\mathcal{A}_{lm}(\omega)}\right)
  \Bigg)
  ~.
\end{eqnarray}
Because the $\mathcal{H}_{lm}$ is proportional to the spherical harmonics $Y_{lm}$, there are symmetries of $\mathcal{H}_{lm}$ for multipole $m$, namely, $\mathcal{H}_{lm}=(-1)^m\mathcal{H}_{l,-m}$.
Based on Eqs.~(\ref{43a}), the solution of $\delta p^0/p^0$ can be formally given by
\begin{eqnarray}
  \frac{\delta p^0}{p^0} & = & \sum_{l = 2}^{\infty} \sum_{m = - l}^l \int
  \frac{{\rm d} \omega}{2 \pi} \mathcal{A}_{l  m} ({\omega}) \int_{s_0}^s
  {\rm d} s\mathcal{H}_{l  m} (\tau (s), \rho (s), \theta (s), \phi (s)
  ; \bar{\kappa}, \bar{\lambda})) e^{-i \omega r_h \tau} ~ . \label{46a}
\end{eqnarray}
Using Eq.~(\ref{18}), (\ref{27}), (\ref{34}), (\ref{48}) and (\ref{46a}), the redshift fluctuations of light take the form of
\begin{eqnarray}
  \frac{\delta z}{1+z} & = & \sum_{l = 2}^{\infty} \sum_{m = - l}^l \int
  \frac{{\rm d} \omega}{2 \pi} h_{l  m}^{\rm{GW}} (\omega) \mathcal{R}_{l
    m} (\omega; \kappa, \varphi) ~,  \label{46}
\end{eqnarray}
where the dimensionless quantity $\mathcal{R}_{l  m} (\omega, \bar{\lambda})$ is the response function
for the VLBI,
\begin{eqnarray}
  \mathcal{R}_{l  m} (\omega;\kappa, \varphi) & = & 4 \sqrt{\pi \frac{(l - 2)
      !}{(l + 2) !}} \int_{s_{\rm{emt}}}^{s_{\rm{obs}}} {\rm d}
  s\mathcal{H}_{l  m} (\tau (s), \rho (s), \theta (s), \phi (s) ; \kappa, -
  \sqrt{\kappa} \sin \theta_o \sin \varphi)e^{-i \omega r_h \tau} ~. \label{47}
\end{eqnarray}
Eq.~(\ref{46}) can formulate the redshift fluctuations of every single light ray that reaches the image plane. As shown in Eq.~(\ref{47}), the redshift fluctuations could be accumulated along trajectories of light rays from emission regions to the image plane.   Based on the linear perturbation theory, the redshift fluctuation is shown to be a linear response to the gravitational fluctuations. 

\subsection{Correlations and overlap reduction functions on the photon ring}

The intensity fluctuations in Eq.~(\ref{0}) are closely related to emission sources around a black hole \cite{Gralla:2019xty,Hadar:2020fda}. Here, to theoretically study the intensity fluctuations as a result of gravitational fluctuations, our primary interest lies in the geometry of space-time, namely, focus on the behavior of light from the vicinity of photon sphere. These light rays can produce the photon ring on the images, which is demonstrated to be insensitive to emission sources \cite{Johannsen:2015hib,Narayan:2019imo,Gralla:2020srx,Gralla:2020yvo,Bronzwaer:2021lzo}.

For light rays $a$ and $b$ shown in Fig.~\ref{F1}, the correlations of redshift fluctuations can be given by 
\begin{eqnarray}
  {\rm{Corr}}_{z} (\varphi_{a b}) & \equiv & \int_0^{2 \pi} \frac{{\rm d}
    \bar{\varphi}}{2 \pi} \left\langle \left. \frac{\delta z}{1+z}
  \right|_{\bar{\varphi}} \left. \frac{\delta z}{1+z} \right|_{\bar{\varphi} +
  \varphi_{ab}} \right\rangle \nonumber\\
  & = & \sum_{l, m, l', m'} \int \frac{{\rm d} \bar{\varphi}}{2 \pi} \int
  \frac{{\rm d} \omega}{2 \pi} \int \frac{{\rm d} \omega'}{2 \pi} \langle h^{\rm{GW}}_{l
      m} (\omega) h^{\rm{GW}}_{l' m'} (\omega') \rangle \mathcal{R}_{l
    m} (\omega; \kappa_{\rm ps}, \bar{\varphi}) \mathcal{R}_{l'  m'}^{*} (\omega'; \kappa_{\rm ps},
  \bar{\varphi} + \varphi_{ab}) \nonumber\\
  & = & \sum_{l, m} \int \frac{{\rm d} \omega}{2 \pi} P_{l  m} (\omega)
  \int_0^{2 \pi} \frac{{\rm d} \bar{\varphi}}{2 \pi} \mathcal{R}_{l  m}
  (\omega; \kappa_{\rm ps}, \bar{\varphi}) \mathcal{R}_{l  m}^{*} (\omega; \kappa_{\rm ps}, \bar{\varphi}
  + \varphi_{ab}) ~,  \label{51}
\end{eqnarray}
where the polar angles of light rays $a$ and $b$ on image plane are $\varphi_a=\bar{\varphi}$ and $\varphi_b=\bar{\varphi}+\varphi_{ab}$, respectively. To restrict the correlations on the photon ring, we let $\kappa =\kappa_{\rm ps}$. Since the $h^{\rm{GW}}_{l  m} (\omega)$ is
independent of polar angle $\varphi$, the radiated power spectrum $P_{lm}(\omega)$ can be separated from the response function $\mathcal{R}_{lm}(\omega,\varphi)$ in the correlations. Following the approaches used in the field of gravitational wave detectors \cite{Romano:2016dpx}, the overlap reduction function can be read from Eq.~(\ref{51}), namely
\begin{eqnarray}
  \Gamma_{l  m} (\omega, \varphi) & = & \int_0^{2 \pi} \frac{{\rm d}
    \bar{\varphi}}{2 \pi} \mathcal{R}_{l  m} (\omega; \kappa_{\rm ps}, \bar{\varphi})
  \mathcal{R}_{l  m}^{*} (\omega; \kappa_{\rm ps}, \bar{\varphi} + \varphi) ~ . \label{52}
\end{eqnarray}
%It corresponds to $\varphi$-dependent parts of the correlations for given frequency.
In principle, the overlap reduction functions can reflect characteristic signature of the gravitational fluctuations for given black hole background. For example, in Minkowski background, pulsar timing arrays as gravitational wave detectors can provide characteristic signature described by the overlap reduction function, which is known as Hellings-Downs curve \cite{Hellings:1983fr}.% used to distinguish whether the detected stochastic background is the tensor mode.

\begin{figure}
  \includegraphics[width=1\linewidth]{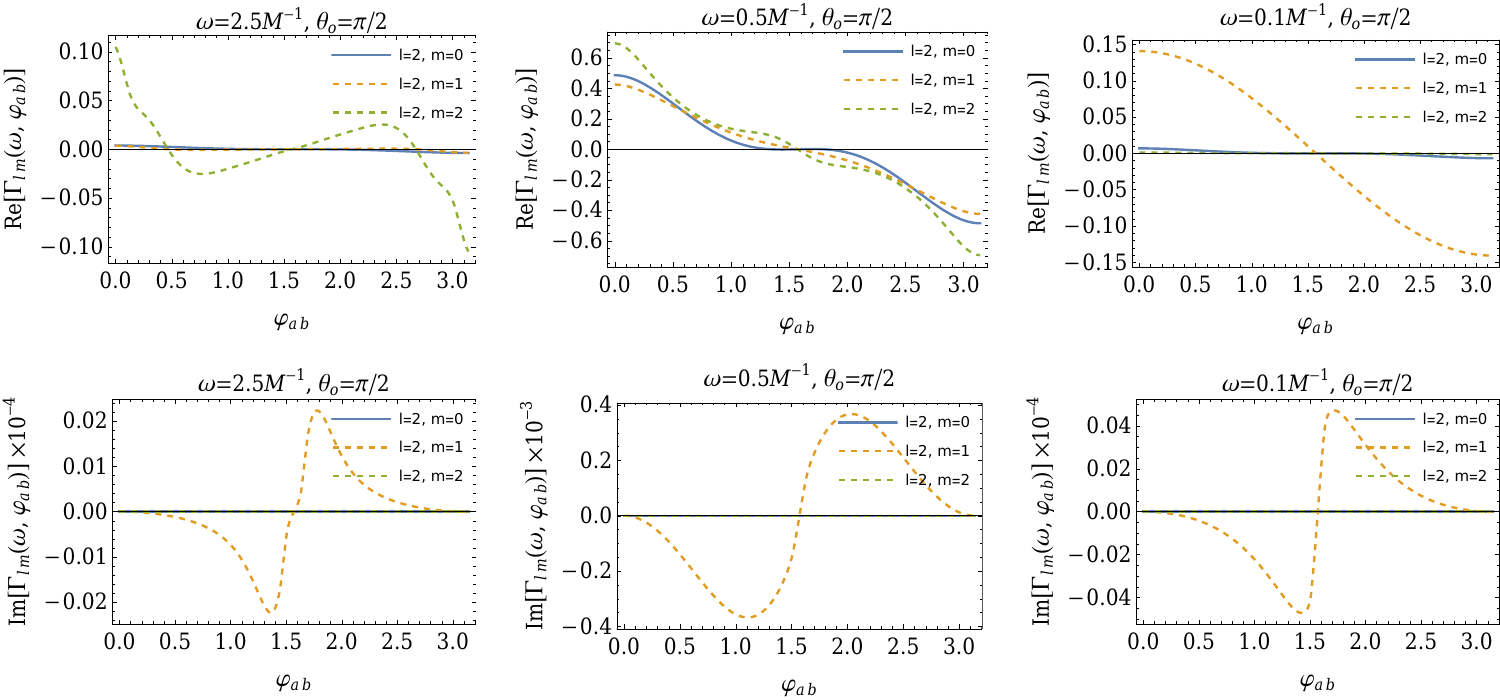}
  \caption{Overlap reduction function as functions of angle
    $\varphi_{ab}$ for selected multipole $m$ and frequency $\omega$ with $l=2$ and $\theta_{\rm o}=\pi/2$.\label{F2}}
\end{figure}

\begin{figure}
  \includegraphics[width=1\linewidth]{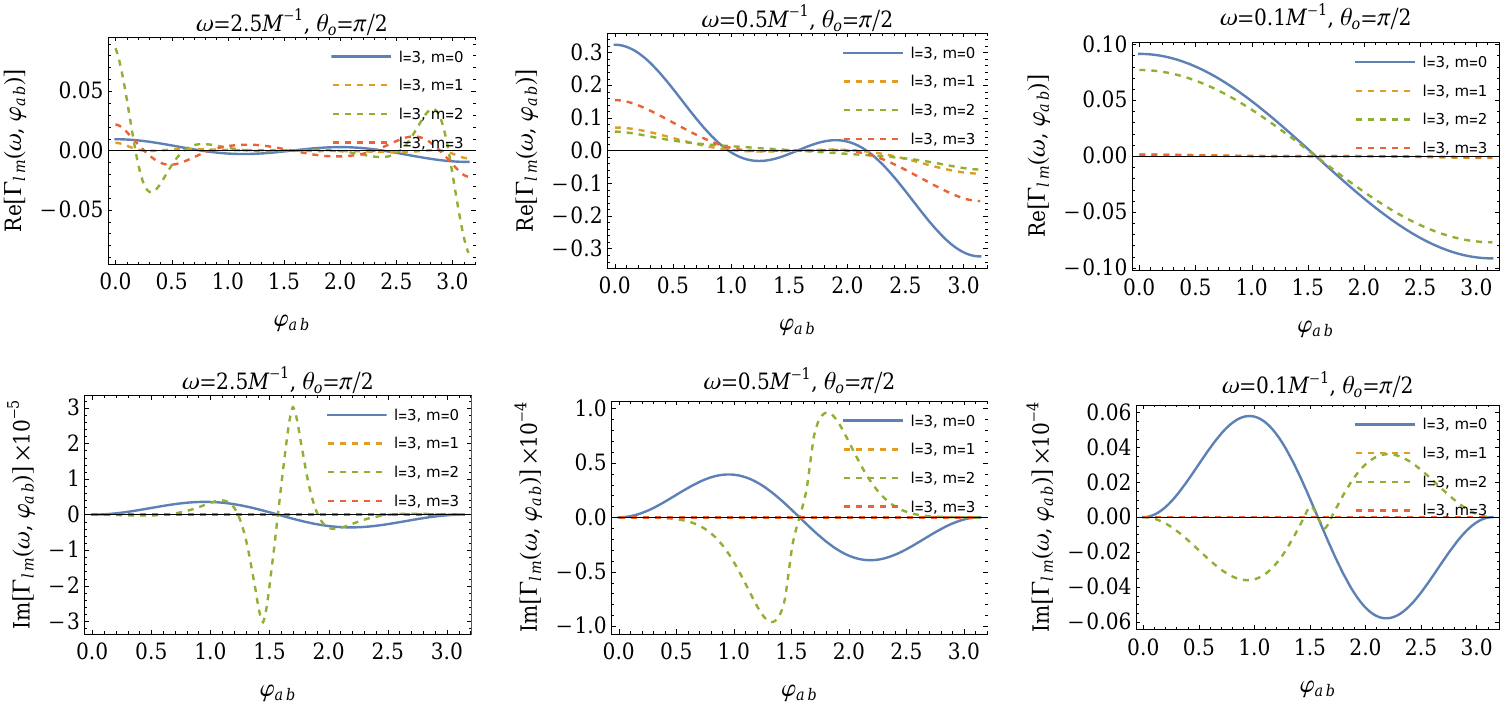}
  \caption{Overlap reduction function as functions of angle
    $\varphi_{ab}$ for selected multipole $m$ and frequency $\omega$ with $l=3$ and $\theta_{\rm o}=\pi/2$.\label{F3}}
\end{figure}

\begin{figure}
  \includegraphics[width=1\linewidth]{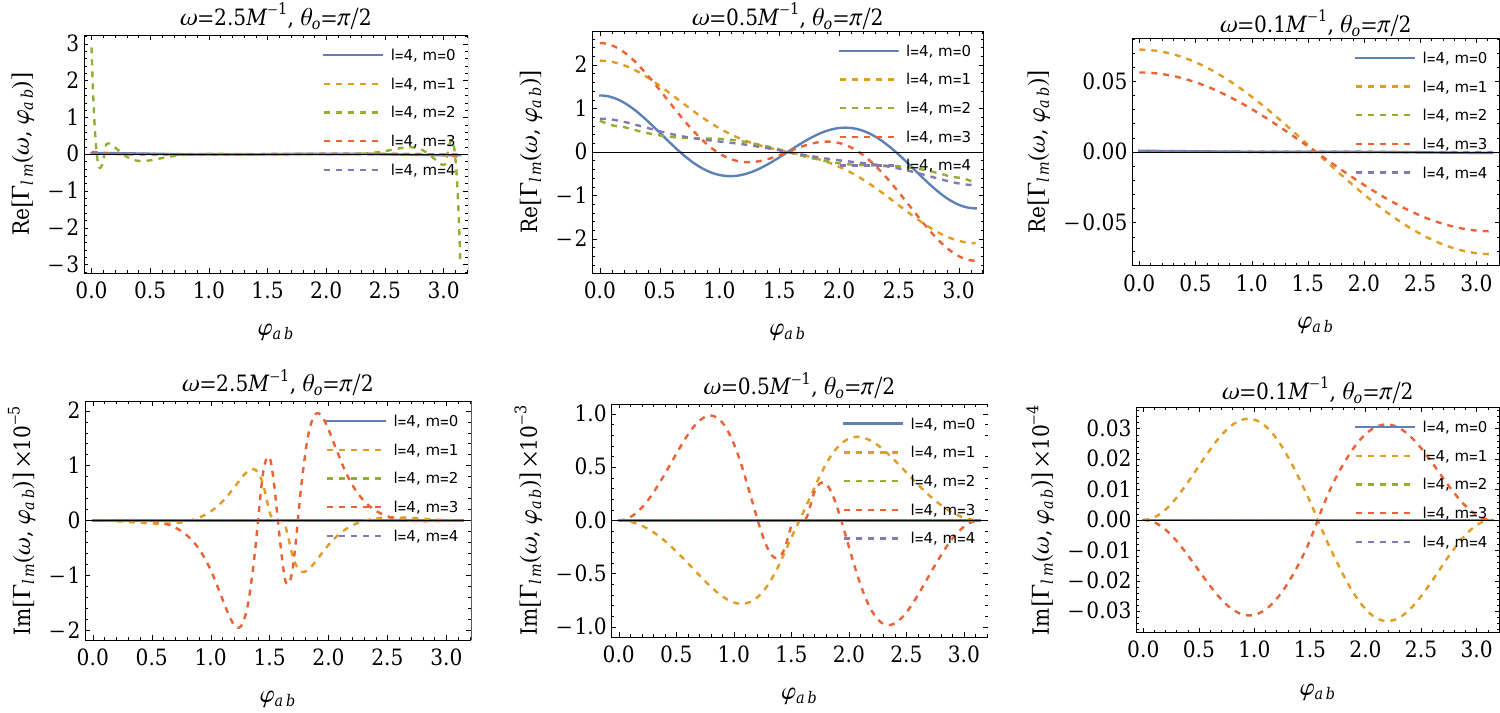}
  \caption{Overlap reduction functions as functions of angle
    $\varphi_{ab}$ for selected multipole $m$ and frequency $\omega$ with $l=4$ and $\theta_{\rm o}=\pi/2$.\label{F4}}
\end{figure}

In Figs.~\ref{F2}, \ref{F3} and \ref{F4}, we show the overlap reductions functions as functions of angle
$\varphi_{ab}$ for selected multipole $l$, $m$.
For the light rays getting close to each other, $\varphi \rightarrow 0$, the real
parts of the overlap reduction functions tend to be the larger, while the
imaginary parts tend to be zero. In the case of $\varphi = \pi / 2$, %namely, for the position vectors of light rays $a$ and $b$ on the image plane that are orthogonal to each other,
there is not correlation for these light rays. We consider the cases of  $\omega = 0.1 M^{-1}$, $0.5 M^{-1}$ and $2.5 M^{-1}$ in the plots representing low frequency, medium frequency and high
frequency of the gravitational fluctuations, respectively. %It is found that the overlap reduction function tends to be much larger for the low-frequency case. In other words, via observing the light rays from a black hole, the response of the detector tends to be sensitive to the low frequency gravitational waves.
For high-frequency gravitational fluctuations, the cases of $m=2$ dominate the values of real parts of overlap reduction functions, while for the low-frequency modes, the cases of $l+m=3,5,7$ tend to give a larger value of real parts of overlap reduction functions. Besides, due to the factors $e^{-i\omega r_{\rm h}\tau}$ in Eq.~(\ref{47}), the imaginary parts of overlap reduction functions are inevitable. It is  expected to be canceled out after integrating $\omega$ in the correlations ${\rm Corr}_z(\varphi_{ab})$.

\begin{figure}
  \includegraphics[width=1\linewidth]{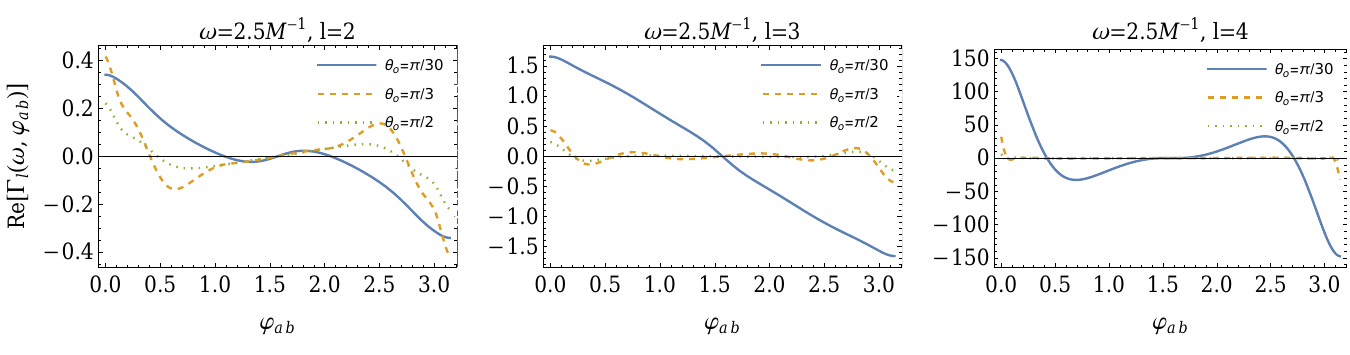}
  \caption{The $\Gamma_l (\omega,\varphi_{ab})$ as function of $\varphi_{ab}$ for selected $\theta_{\rm o}$  and multipole $l$ with $\omega=2.5M^{-1}$\label{F5}}
\end{figure}

\begin{figure}
  \includegraphics[width=1\linewidth]{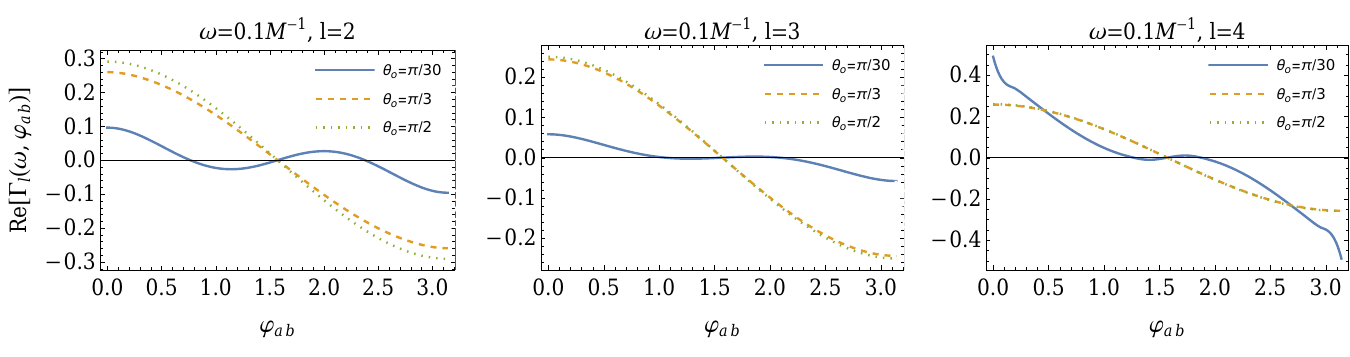}
  \caption{The $\Gamma_l (\omega,\varphi_{ab})$ as function of $\varphi_{ab}$ for selected $\theta_{\rm o}$  and multipole $l$ with $\omega=0.1M^{-1}$\label{F6}}
\end{figure}

\begin{figure}
  \includegraphics[width=1\linewidth]{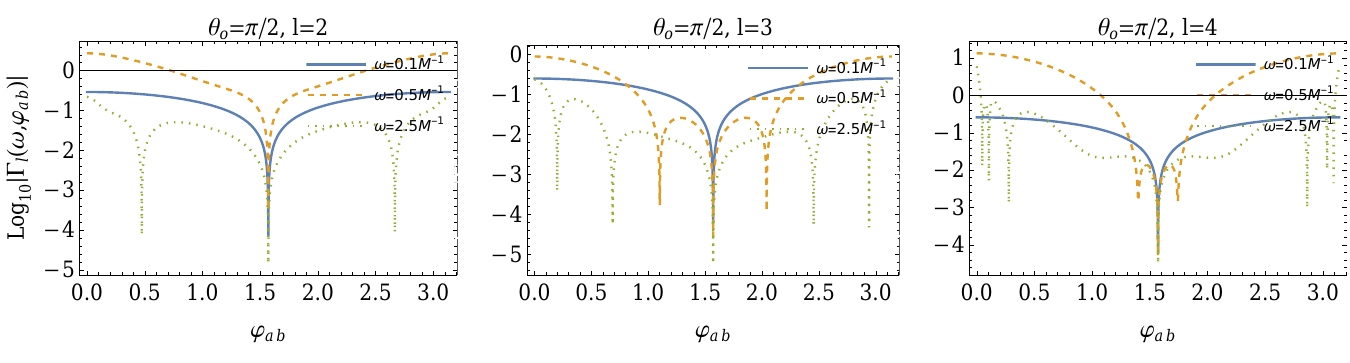}
  \caption{$\Gamma_l (\omega,\varphi_{ab})$ as function of $\varphi_{ab}$ for selected frequency $\omega$ and multipole $l$. \label{F7}}
\end{figure}

It is noted that without given expression of $P_{l  m} (\omega)$, one can not obtain angular
correlations of the ${\rm{Corr}}_{\delta z / z} (\varphi)$ with the sum of $l$ and $m$. Here, we further restrict our calculations to the  cases that the $P_{l  m}
  (\omega)$ is free of multipole $m$. This is allowed in principle, because the evolution equation in Eq.~(\ref{6}) are independent of the $m$. In this case, the
radiated power spectrum can be expressed as $P_{l  m} (\omega)
  \equiv P_l (\omega)$, which results in the two-point correlation taking the form of
\begin{eqnarray}
  {\rm{Corr}}_{\delta z / z} (\varphi) & = & \sum_{l = 2}^{\infty} \int
  \frac{{\rm d} \omega}{2 \pi} P_l (\omega) \sum_{m = - l}^l \Gamma_{l  m} (\omega,
  \varphi) ~ ,
\end{eqnarray}
where the overlap reduction function in multipole $l$ is given by
\begin{eqnarray}
  \Gamma_l (\omega, \varphi) & = & \sum_{m = - l}^l \Gamma_{l  m} (\omega,
  \varphi) ~ .
\end{eqnarray}
In Figs.~\ref{F5} and \ref{F6}, we present the $\Gamma_l (\omega,\varphi_{ab})$ as function of $\varphi_{ab}$ for selected $\theta_{\rm o}$ in low-frequency and high-frequency cases, respectively. It shows that the overlap reduction functions depend on the inclination angle $\theta_{\rm o}$. It is reasonable, since there is not spherical symmetry for the gravitational fluctuations with the multipole $l\geq2$. In the high frequency cases $\omega=2.5M^{-1}$, the values of overlap reduction functions tend to be larger for the cases $\theta_{\rm o}\rightarrow 0$. In the low frequency cases $\omega=0.1M^{-1}$, the shapes of overlap reduction functions seem to be the same order of magnitude for different $l$. %Further study might focus on whether the same shape of the overlap reduction functions is a universal prosperity for the correlations.
In Fig.~\ref{F7}, we presents the $\Gamma_l (\omega,\varphi_{ab})$ as function of $\varphi_{ab}$ for selected frequency $\omega$. It is found that the amplitude of the overlap reduction functions tends to be the largest in the case of $\omega=0.5M^{-1}$. It might be understood as parametric resonance.
For different $\omega$, it shows that there seems to be no order-of-magnitude difference in the values of $\Gamma_l (\omega,\varphi_{ab})$ for lower multipole $l$. 

%The two-point correlations of redshift fluctuations calculated here are the equal-time correlations. Namely, we consider the correlations of two light rays that experience the same time in the Schwarzschild background. To calculate temporal correlations, one should consider critical behavior of light rays near the photon sphere in present of black hole perturbations. It seems not trivial, because the perturbed metric turns to be time-dependent on the photon sphere.
Compared with pioneers's study that attributes the intensity fluctuation to the stochastic emission sources in Fig.~1 of Ref.~\cite{Hadar:2020fda}, here, the two-point correlations as functions of polar angle $\varphi_{ab}$ are shown to be different in shapes. At least, the correlations in present paper i) vanish at the point of $\varphi_{ab}=\pi/2$, and ii) have opposite values at $\theta_{ab}=0$ and $\theta_{ab}=\pi$. We can conclude that the correlations can distinguish the emission sources and black hole geometries in the stochastic regime in principle.

\begin{figure}
  \includegraphics[width=1\linewidth]{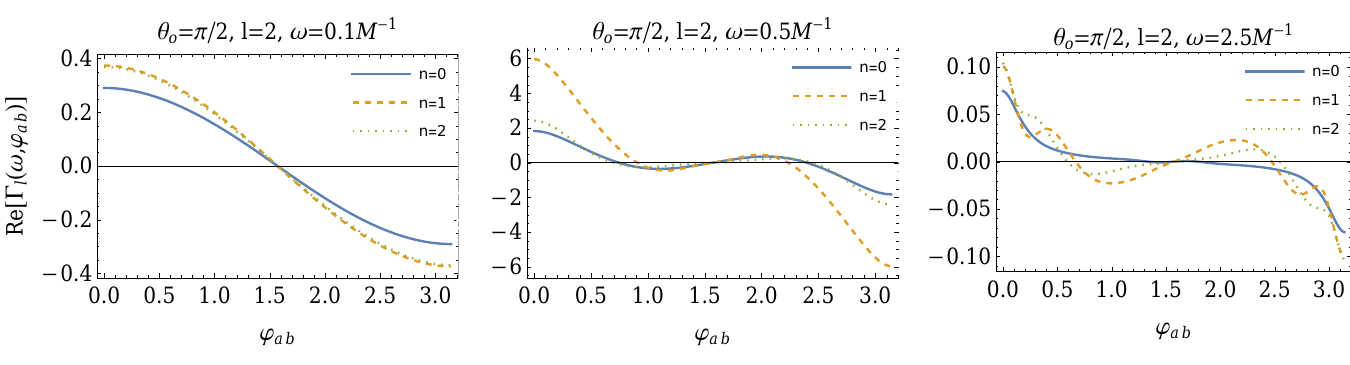}
  \caption{$\Gamma_2 (\omega,\varphi_{ab})$ as function of $\varphi_{ab}$ for selected frequency $\omega$ and lensing band $n$\label{F9}}
\end{figure}

\begin{figure}
  \includegraphics[width=1\linewidth]{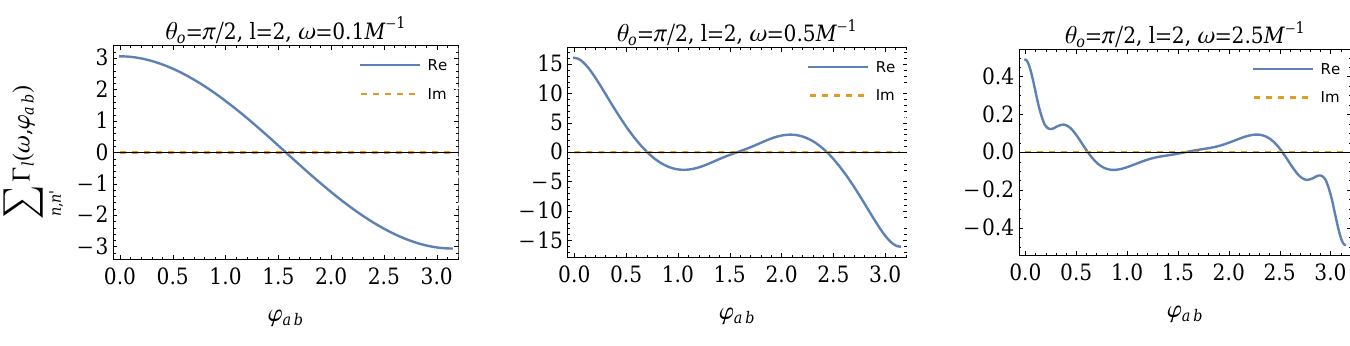}
  \caption{$\sum_{n=0}^{2}\sum_{n'=0}^{2}\Gamma_2 (\omega,\varphi_{ab})$ as function of $\varphi_{ab}$   for selected frequency $\omega$.\label{F10}}
\end{figure}

Due to existence of self-similar structure of photon ring, the light rays from different order of lensing bands might be imaged to be the photon sub-rings \cite{Johnson:2019ljv}, which are identified with half-orbit number $n$ of the light rays. In our study, the  light rays with different $n$ propagating in a perturbed space-time  experience different time in the vicinity of photon sphere. Therefore, it could lead to different imprints on the overlap reduction functions. As shown in Eq.~(\ref{51}), it can be derived from correlations of redshift fluctuations of two light rays $a$ and $b$ for given lensing bands $n$. To be specific, in Fig.~\ref{F9}, the overlap reduction function $\Gamma_l(\omega,\varphi_{a b})$ as function of angle $\varphi_{a b}$ for different $n$ is presented. In high-frequency case $\omega=2.5M^{-1}$, it indicates more micro-structure for the case $n=2$, as more peaks and valleys are shown in the angular correlation curves. In the low-frequency case $\omega=0.1M^{-1}$, there are only difference on the amplitude of the $\Gamma_l(\omega,\varphi_{a b})$. In Fig.~\ref{F10}, we also consider the situations that the light rays $a$ and $b$ are from different lensing bands $n$ and $n'$, respectively. %By taking account of the different photon sub-rings, the shapes of $\Gamma_l(\omega,\varphi_{a b})$ exhibit the greatest changes in high-frequency cases, and 
Because the amplitudes of the $\Gamma_l(\omega,\varphi_{a b})$ are close to be order one for different $n$, it implies that the total amplitude of overlap reduction functions $\Sigma_{n,n'}\Gamma_l(\omega,\varphi_{a b})$ must get larger as a result of the accumulation of sub-ring structure. 
Besides, it is also expected that the angular correlation curves might approach some limit as the increase of the order of the lensing bands. However, because the gravitational fluctuations are oscillating in the vicinity of the photon sphere, this picture is no longer obvious, and we can not provide comprehensive results limited by the numerical approaches. To further investigate this topic, our future studies might employ approximation or analytical methods.

%For spherical black hole images, the intensity correlation from stochastic emission source is shown to be trivial for light rays $a$ and $b$ from the same lensing bands \cite{Hadar:2020fda}. 


\section{Observability of gravitational fluctuations with black hole images  \label{IVa}}

As shown in Eq.~(\ref{0}), the observed intensity fluctuations can result from the stochastic emission sources or redshift fluctuations of light.
%\begin{eqnarray}
% I_{\rm{obs}} (t, \rho, \varphi) & = & \left( 1 + z \left( \bm{x}_o,
% \bm{x}_s \right) \right)^3 I_{\rm{emt}} \left( t, \bm{x}_s (\rho,
% \varphi) \right) ~, 
%\end{eqnarray}
In Sec.~\ref{III}, it shows that the redshift fluctuations exhibit a linear response to the gravitational fluctuations.  In order to detect gravitational fluctuations, the $\delta z / (1 + z)$  should be considered as the signal, and the $\delta I_{\rm{emt}} / \bar{I}_{\rm{emt}}$ is considered as the noise. In a more comprehensive derivation, where we have $I_{\rm obs} = \bar{I}_{\rm obs}+ \delta I_{\rm obs}-\delta I_{\rm ins}$ and $I_{\rm emt} = \bar{I}_{\rm obs}+ \delta I_{\rm emt}$, one can obtain
\begin{eqnarray}
  \underset{\rm{output}}{\frac{\delta I_{\rm{obs}}}{\bar{I}_{\rm{obs}}}}
  & = & \underset{\rm{noise}}{\frac{\delta
  I_{\rm{ins}}}{\bar{I}_{\rm{obs}}}+ \frac{\delta
  I_{\rm{emt}}}{\bar{I}_{\rm{emt}}}} + \underset{\rm{signal}}{3 \left(
    \frac{\delta z}{1 + z} \right)} ~, \label{52}
\end{eqnarray}
%where $\bm{x}_o$ and $\bm{x}_s$ are the location of observers, and emission source, respectively. 
where the $\delta I_{\rm ins}$ is zero-mean intensity fluctuations from instrumental noise, the $\bar{I}_{\rm{obs}}$ is time-averaged ray-traced image of
 emission intensity $\bar{I}_{\rm{emt}}$, and in a similar manner, the $\delta{I}_{\rm{obs}}$ is  ray-traced image of the  $\delta{I}_{\rm{emt}}$. In this context, the fluctuations $\delta I_{\rm ins}$, $\delta I_{\rm emt}$, $\delta I_{\rm obs}$, and $\delta z$ are $t$-dependent, and the background quantities $ \bar{I}_{\rm emt}$, $\bar{I}_{\rm obs}$, and $ z$ are considered to be $t$-independent. 

Suppose that the intensity and redshift fluctuations are mean-subtracted Gaussian stochastic variables. For a single point on the images, the signal-to-noise ratio (SNR) can be given by \cite{10.5555/559989}
\begin{eqnarray}
  \rm{SNR}^2 & = & T \int^{\frac{\pi}{\delta t}}_0  \left( \frac{P_{\Delta z} (\omega)}{P_{\Delta I} (\omega)} \right)^2 \frac{ {\rm d} \omega}{2
  \pi}
  ~, \label{53}
\end{eqnarray}
where the $T$ is observation duration, the $\delta t$ is time sampling interval
(cadence), the spectral density of redshift fluctuations was obtained in Sec.~\ref{III}, namely,
\begin{eqnarray}
  P_{\Delta z} (\omega) & = & 18 \pi \frac{ {\rm d}}{ {\rm d} \omega} {\rm{Corr} }(0) = 18
  \pi \sum_l P_{l  } (\omega) \Gamma_{l  }(\omega,0) ~, \label{54}
\end{eqnarray}
and the $P_{\Delta I} (\omega)$ is spectral density of the intensity fluctuations resulting from stochastic emission sources and instrumental noise. Because there would be no correlation between emission sources and VLBI instruments, the $P_{\Delta I} (\omega)$ can be expressed as  $P_{\Delta I} (\omega)= P_{\Delta I,\rm s} (\omega) +P_{\Delta I,\rm n} (\omega)$, where $P_{\Delta I,\rm s} (\omega)$ and $P_{\Delta I,\rm n} (\omega)$ are spectrum of observed stochastic emission intensity and spectrum of instrumental noise, respectively.
In the following, we will derive a phenomenological expression of $P_{\Delta I} (\omega)$ based on pioneers' simulations \cite{EventHorizonTelescope:2022ago,EventHorizonTelescope:2022okn}, and then provide an order-of-magnitude estimate for the $\text{SNR}$ in Eq.~(\ref{53}).

\subsection{Spectral density of the noise }
The observables of VLBI array of telescopes are visibilities defined on the image plane. Theoretically, it can be derived from the intensity, namely,
\begin{eqnarray}
  V \left( t, \bm{u} \right) & = & \int  {\rm d}^2 y   I_{\rm{obs}}
  \left( t, \bm{y} \right) e^{2 \pi i   u \cdot y} ~,
\end{eqnarray}
where the $\bm u$ is vector baseline, and $\bm{y}=(\alpha,\beta)$ is the two-dimensional Cartesian coordinate on the plane of images. 
The $t$-dependence of visibility or intensity might be originated from evolving emission sources according to underlying astrophysical processes \cite{EventHorizonTelescope:2022ago,EventHorizonTelescope:2022okn} or simply instrumental noise \cite{Wielgus:2021peu}.

%As previously stated, the intensity distributions can be decomposed in the form of
%$%\begin{eqnarray}
%  I_{\rm{emt}} \left( t,   \bm{x} \right)  =
%  \bar{I}_{\rm{emt}} \left( \bm{x} \right) + \delta I_{\rm{emt}}
%  \left( t,   \bm{x} \right)
%$. %\end{eqnarray}

%\subsubsection{Variability of emission sources}

For intensity fluctuations originated from emission sources, one can derive the power spectra  $P_{\Delta I, \rm s} (\omega)$ from the $\delta I_{\rm{emt}} \left( t,   \bm{x} \right)$.
Employing ray-tracing, the images of  $\delta I_{\rm{emt}} \left( t,   \bm{x} \right)$ can be obtained via $\delta I_{\rm{obs,s}} \left(
 t, \bm{y}\right) \equiv (1 + z)^3 \delta I_{\rm{emt}} \left(t, \bm{x}_s
  \left( \bm{y} \right) \right)$, where $\bm{x}_s$ is the location of emission source. And one can introduce visibility $\delta V
  \left(t, \bm{u}\right) \equiv \int  {\rm d}^2 y \delta
  I_{\rm{obs,s}} \left(t, \bm{y}\right) e^{2 \pi i   u \cdot y}$
and $\bar{V} \left( \bm{u} \right) \equiv \int  {\rm d}^2 y
  \bar{I}_{\rm{obs}} \left( \bm{y} \right) e^{2 \pi i   u \cdot
      y}$, such that $V \left(t, \bm{u}\right) = \bar{V} \left( \bm{u}
  \right) + \delta V \left(t, \bm{u}\right)$.
To perform statistical measures of the visibilities in temporal Fourier space, we ansate
\begin{eqnarray}
  \left\langle \delta \tilde{V} \left( \omega, \bm{u}\right) \delta
  \tilde{V}^{*} \left(w', \bm{u} \right) \right\rangle & = & 2
  \pi \delta  (\omega - w') P_{\delta V} \left( \omega, \bm{u}\right) ~,
\end{eqnarray}
where $\delta \tilde{V} \left( \omega, \bm{u}\right)$ is the temporal Fourier
mode of the $\delta V \left(t, \bm{u}\right)$, namely,
$
  \delta V \left(t, \bm{u}\right)  =  \int \frac{ {\rm d} \omega}{2 \pi}
  \delta \tilde{V} \left( \omega, \bm{u}\right) e^{i  \omega  t}
$, and the $P_{\delta V}(\omega,\bm u)$ is power spectrum of the stochastic variable $\delta V$.

The variance of the $\delta V \left(t, \bm{u}\right)$ can quantify amplitude of the variability. It was obtained by Georgiev et al. based on GRHMD simulations \cite{EventHorizonTelescope:2022okn}, namely,
\begin{eqnarray}
  \langle P_{ \mathcal{T}} \rangle & \equiv & \left\langle \delta V \left( t,\bm{u}
  \right) \delta V \left(t, \bm{u}\right) \right\rangle_{ \mathcal{T}} = \int
  \frac{ {\rm d} \omega}{2 \pi} P_{\delta V} \left( \omega, \bm{u}\right) W_{ \mathcal{T}} (\omega)
  ~, \label{59}
\end{eqnarray}
where the $W_{ \mathcal{T}} (\omega)[ = (1 - e^{- 2 \pi   w^2 \mathcal{T}^2})^2]$ is window function derived from Gaussian smoothing.
Here, one can obtain the $P_{\delta V} \left( \omega, \bm{u}\right)$, due to the fact that \cite{EventHorizonTelescope:2022okn}
\begin{eqnarray}
  \langle P_{ \mathcal{T}} \rangle_{T = (2 \pi \omega)^{- 1}} & \simeq & \frac{\Delta \omega}{2 \pi}
  P_{\delta V} \left( \omega, \bm{u}\right) ~, \label{60}
\end{eqnarray}
From Refs.~\cite{EventHorizonTelescope:2022ago,EventHorizonTelescope:2022okn}, the $\langle P_{ \mathcal{T}} \rangle_{\mathcal{T} = (2 \pi \omega)^{- 1}}$  represent a red-noise process for $\omega \gtrsim 0.01 M^{- 1}$. Namely, we have a phenomenological expression
$\langle P_{ \mathcal{T}} \rangle_{\mathcal{T} = (2 \pi \omega)^{- 1}} = A  (\omega / \omega_{\ast})^{n_s}$, where
$\omega_{\ast} = M^{-1}$, $n_s \simeq - 2$, and $A (0) \simeq 6 \times 10^{- 6}
  \rm{Jy}^2$. Assuming the $P_{\delta V}\left( \omega, \bm{u}\right)$ in the form of a power-law spectrum, one can estimate $\Delta\omega\simeq 0.1
  w$ based on Eq.~(\ref{59}). Associated above results with Eq.~(\ref{60}), we thus obtain
\begin{eqnarray}
  P_{\delta V} \left( \omega, \bm{u}\right) & \simeq & 20 \pi A \left(
  \bm{u} \right) \omega_{\ast}^{- 1} \left( \frac{\omega}{\omega_{\ast}} \right)^{n_s -
    1} ~.
\end{eqnarray}

To further obtain the $P_{\Delta I,\rm s}$, we introduce 
spectrum $P_{\delta I / I}(\omega, {\bm y})$ with respect to the intensity fluctuations $\delta I_{\rm{obs,\rm s}} /
  I_{\rm{obs}}$. Since $\delta V$ is the Fourier modes of $\delta
  I_{\rm{obs}}$, one can derive $P_{\delta I / I}$ via
\begin{eqnarray}
  \int  {\rm d}^2 y \left\{ \bar{I}_{\rm{obs}} \left( \bm{y} \right)
  P_{\delta I / I} \left(\omega , \bm{y}\right) \right\} & = & \int  {\rm d}^2 u
  P_{\delta V} \left( \omega, \bm{u}\right) \simeq P_{\delta V} (0,
  \omega) \int_{\mathcal{S}}  {\rm d}^2 u ~, \label{62}
\end{eqnarray}
where $\mathcal{S}$ is a finite region determined by $|{\bm u}|\lesssim 10 \text{G}\lambda$, approximately. The second equal sign is obtained based on the shape of $P_{\delta V} \left(
  \omega, \bm{u}\right)$ for different baseline $\bm u$ given in Ref.~\cite{EventHorizonTelescope:2022okn}.  One can give an
order-of-magnitude estimate by providing intensity
 $\bar{I}_{\rm{obs}} (\rho, \varphi)$ in the near region of the photon rings, because it was shown that the photon ring contains
20\% of the total image flux density \cite{Johnson:2019ljv}. Namely, on the left hand side of Eq.~(\ref{62}), one can let the domain of integration be $\rho \in
  \left( \rho_c - \frac{1}{2} \delta \rho, \rho_c + \frac{1}{2} \delta \rho
  \right)$, where $\rho_c$ is the radius of critical curves. %Additionally, for spherical black hole in this study, we can further simplify the derivations, since the $\bar{I}_{\rm{obs}}$ near the photon rings is independent of $\varphi$.
 %taking into account above considerations, 
And thus, it can  be evaluated in the form of %the left hand side of Eq.~(\ref{62}) can be evaluated in the form of
\begin{eqnarray}
  \int  {\rm d}^2 y \left\{ \bar{I}_{\rm{obs}} \left( \bm{y} \right)
  P_{\delta I / I} \left( \omega, \bm{y}\right) \right\} & \simeq & \pi
  \rho_c^2 \delta \rho \bar{I}_{\rm{obs}} (\rho_c) \int \frac{ {\rm d}
    \varphi}{2 \pi} P_{\delta I / I} (\omega, \rho_c, \varphi) ~, \label{63}
\end{eqnarray}
where the $\bar{I}_{\rm{obs}}$ in near region of the photon ring is independent of $\varphi$ for spherical black holes. 
In accordance with the definition of $P_{\Delta
      z} (\omega)$ shown in Eqs.~(\ref{51}) and (\ref{54}), here, the 
$P_{\Delta I,\rm s} (\omega)$ should be given with the integration of $\varphi$ on the photon ring. And using Eqs.~(\ref{63}) and (\ref{62}), we obtain the expression of $P_{\Delta I,\rm s} (\omega)$, namely,
\begin{eqnarray}
  P_{\Delta I,\rm s} (\omega) & = & \int \frac{ {\rm d} \varphi}{2 \pi} P_{\delta I / I}
  (\rho_c, \varphi, \omega) \nonumber\\
  & \simeq &   \frac{P_{\delta V} (0, \omega)}{\pi \rho_c^2 \delta \rho
  \bar{I}_{\rm{obs}} (\rho_c)} \int_{\mathcal{S}}  {\rm d}^2 u \simeq
  \frac{P_{\delta V} (0, \omega)}{\bar{V} (0)^2} \nonumber\\
  & = & 20 \pi \frac{A \left( 0 \right)}{\bar{V}_0 (0)^2}
  \omega_{\ast}^{- 1} \left( \frac{\omega}{\omega_{\ast}} \right)^{n_s - 1} ~.
\end{eqnarray}
In third equal sign, we have used $\bar{V} (0) = \int  {\rm d}^2 y
  \bar{I} \left( \bm{y} \right) \simeq \pi \rho_c^2 \delta \rho \bar{I}
  (\rho_c)$ and thereby $\int_{\mathcal{S}}  {\rm d}^2 u \simeq \pi \rho_c^2
  \delta \rho$. By providing $\bar{V} (0) \simeq 2.5 \rm{Jy}$ from the GRHMD simulations in Ref~\cite{EventHorizonTelescope:2022okn}, we finally obtain
\begin{eqnarray}
  P_{\Delta I,\rm s} (\omega) & \simeq & 6 \times 10^{- 5} \omega_{\ast}^{- 1} \left(
  \frac{\omega}{\omega_{\ast}} \right)^{- 3} ~, \text{ for } w>0.01w_*~. \label{65}
\end{eqnarray}
The spectral density describing the variability of intensity from stochastic emission sources will be considered as the one of noise spectra in the gravitational fluctuation detection.

%\subsubsection{White noise from instruments}
Besides the intensity fluctuation from emission sources, there would be also uncertainty of observation in a practical instrument, specifically, the $\delta I_{\rm ins}$ shown in Eq.~(\ref{52}). Here, we consider the instrument noise as a white noise. In this context, the $P_{\Delta I, \rm n}$ should be a flat spectrum, namely,
%$
%  \sigma^2 & = & \int_0^{\frac{\pi}{\delta t}} \frac{{\rm d} w}{2 \pi}
%  P_{\Delta I, n} (w) = \frac{P_{\Delta I, n} (w)}{2 \delta t}
%$
%Since $P_N = \rm{const} .$ we have
\begin{eqnarray}
  P_{\Delta I, \rm n}(\omega) & = & 2 \sigma^2 \delta t~, \label{66}
\end{eqnarray}
where the $\delta t$ is cadence, and the $\sigma$ is standard deviation of the $\delta I_{\rm ins}/\bar{I}_{\rm obs}$. From the study on the variability of Sgr A* light curves \cite{Wielgus:2021peu}, we have order-of-magnitude estimates of the deviation, $\sigma\lesssim 0.01$.

\subsection{Characteristic strain of gravitational waves and sensitivities}
 
The gravitational fluctuations being generated near a black hole and then propagating to the distant detectors, known as gravitational waves, are the one of the objectives of observatories, such as LIGO, LISA and PTAs. Because the characteristic strain of gravitational waves is defined at locations of detectors, we  here consider angular-averaged characteristic strain at distance $r$ in the form of  
\begin{eqnarray}
  \int  {\rm d}   \ln  \omega  h_c(\omega,r)^2 & \equiv & \int
  \frac{ {\rm d} \Omega}{4 \pi} \left\langle \delta g_{i   j}^{\rm TT}
  \left( t, r, \widehat{\bm{x}} \right) \delta g_{i   j}^{\rm TT} \left( t, r, \widehat{\bm{x}} \right) \right\rangle \nonumber\\
  & = & \int  {\rm d}   \ln  \omega\left\{ \frac{1}{2 \pi  \omega   r^2} \sum_{l = 2}^{\infty} \sum_{m = - l}^l \frac{(l + 2) !}{(l -
    2) !} P_{\mathcal{A}, l   m} (\omega) \right\}, \hspace{1.5em} \text{for
    large } r \nonumber\\
  & = & \int  {\rm d}   \ln  \omega\left\{ \frac{8}{  r^2}
  \sum_{l = 2}^{\infty} \omega^{- 1} P_{l  } (\omega) \right\}~.
\end{eqnarray}
%where we have performed an averaging of the solid angle for the correlations. 
Thus, the characteristic strain $h_c$ can be given by
\begin{eqnarray}
  h_c(\omega,r)^2 & = & \frac{8}{  r^2} \sum_{l = 2}^{\infty} \omega^{- 1} P_l \label{68}
  (\omega) ~,
\end{eqnarray}
where $h_c$ is a dimensionless quantity, and decays with $r^{-1}$ in the leading order. For given multipole $l$,
one can also introduce characteristic strain $h_{c, l} (\omega,
  r)$, such that $h_c^2 = \sum_l h_{c, l}^2$. 
  
  Based on Eqs.~(\ref{53}), (\ref{54}), (\ref{65}), (\ref{66}) and (\ref{68}), we obtain the $\rm{SNR}^2$ in each bin of frequency centered around $f$ with bandwidth $\Delta \omega/2\pi$, namely,
  \begin{eqnarray}
    \rm{SNR}^2 & = & 1.4 \times 10^{10} (\omega_{\ast} r)^4 T \sum^{\infty}_{l = 2}
    \int_0^{\frac{\pi}{\delta t}} \frac{ {\rm d} \omega}{2 \pi} \left(
    \frac{\frac{\omega}{\omega_{\ast}}}{\left( \frac{\omega}{\omega_{\ast}} \right)^{- 3} + 3.0
    \times 10^4 \sigma^2   \omega_{\ast} \delta t}   h_{c, l} (\omega, r)^2
    \Gamma_{l  } (\omega, 0) \right)^2 \nonumber\\
    & \simeq & \left. 1.4 \times 10^{10} (\omega_{\ast} r)^4 \left( \frac{T \Delta
    \omega}{2 \pi} \right) \left( \frac{\frac{\omega}{\omega_{\ast}}}{\left( \frac{\omega}{\omega_{\ast}}
    \right)^{- 3} + 3.0 \times 10^4 \sigma^2   \omega_{\ast} \delta t}
      \right)^2 \sum^{\infty}_{l = 2} h_{c, l} (\omega, r)^4 \Gamma_{l
     } (\omega, 0)^2 \right|_{\omega = 2 \pi f}~, 
  \end{eqnarray}
where we have $f\lesssim 1/2\delta t$ according to upper bound of the integration.
%\begin{eqnarray}
%  \rm{SNR}^2 & = & T \int_0^{\frac{\pi}{\delta t}} \frac{ {\rm d} \omega}{2 \pi}
%  \left( \frac{18 \pi \times \frac{1}{8}\omega  r^2 \sum_l h_{c, l} (\omega,
%      r)^2 \Gamma_{l  }(\omega,0)}{6 \times 10^{- 5} \omega_{\ast}^{- 1} \left(
%      \frac{\omega}{\omega_{\ast}} \right)^{- 3}} \right)^2 \nonumber\\
%  & = & 10^{10} (\omega_{\ast} r)^4 T \sum^{\infty}_{l = 2}
%  \int_0^{\frac{\pi}{\delta t}} \frac{ {\rm d} \omega}{2 \pi} \left\{ \left(
%  \frac{\omega}{\omega_{\ast}} \right)^8 | \Gamma_l(\omega,0) |^2 h_{c, l}(\omega,r)^4
%  \right\} \nonumber\\ 
%  & \simeq & 2 \times 10^8 (\omega_{\ast} r)^4 (\omega_{\ast} T) \left(
%  \frac{\omega}{\omega_{\ast}} \right)^9 \left. \sum^{\infty}_{l = 2} | \Gamma_l(\omega,%0)
%  |^2 h_{c, l}(\omega,r)^4 \right|_{\omega = \frac{\pi}{\delta t}} ~.
%\end{eqnarray}
%The third equal sign is given based on $\Gamma_l(\omega,0) \sim \mathcal{O} (1)$
%according to the results in Sec.~\ref{III}, and the assumptions of $h_{c, l} \propto  (\omega /  \omega_{\ast})^{n_h}$ for $n_h > - 5$. 
The SNR is proportional to square of the distance to the black holes, because the $h_c$ is defined at the locations on the earth, and the images of VLBI observation consist of the light that can carry information of space-time geometries in near regime of a black hole.

For further illustrations, we obtain the $\rm SNR$ for the supermassive black hole Sgr A* in our galaxy, namely,
%\begin{eqnarray}
%  {\rm{SNR}}^2_{\text{Sgr A*}} & = & 2 \times 10^8 \left(
%  \frac{r_{\text{Sgr A*}}}{2 \times 10^{- 10} {\rm{kpc}}} \right)^4 \left(
%  \frac{T}{7 \times 10^{- 7} {\rm{yr}}} \right) \left( \frac{f}{8 \times 10^{-
%      2} {\rm{Hz}}} \right)^9 \nonumber\\
%  & & \times \left. \sum^{\infty}_{l = 2} | \Gamma_l (f, 0) |^2
%  h_{c, l} \left( f, r_{\text{Sgr A*}} \right)^4 \right|_{f = \frac{1}{2 \delta
%    t}} ~.
%\end{eqnarray}
\begin{eqnarray}
  \rm{SNR}^2_{\text{Sgr*A}} & = & 1.4 \times 10^{10} \left(
  \frac{r_{\text{Sgr*A}}}{2.1 \times 10^{- 10} \rm{kpc}} \right)^4 \left(
  \frac{T}{6.7 \times 10^{- 7} \rm{yr}} \right) N^{\frac{1}{4}} \left(
  \frac{\delta t}{10 \rm s} \right)^{\frac{1}{4}} \nonumber\\
  &  & \times \left( \frac{\frac{f}{7.6 \times
  10^{- 3} \rm{Hz}}}{\left( \frac{f}{7.6 \times 10^{- 3} \rm{Hz}}
  \right)^{- 3} + \left( \frac{\delta t}{6.3 \times 10^{- 4} \rm s} \right)
  \sigma^2} \right)^2   \sum^{\infty}_{l = 2} | \Gamma_l (f, 0) |^2 h_{c, l}
  \left( f, r_{\text{Sgr*A}} \right)^4 ~
  , 
\end{eqnarray}
where we adopt $\Delta \omega \simeq \pi/(N\delta t)$, and the $N$ is the number of bins.
With the known distance to the Sgr A*, $(2.1 \times 10^{- 10} \rm{kpc})^{-
      1} r_{\text{Sgr A*}} \approx 4 \times 10^{10}$, and the criterion $\rm{SNR}^2 \gtrsim 1$, we obtain sensitivity curves in terms of the characteristic strain with $l=2$, namely, 
%\begin{eqnarray}
%  h_{c, l} \left( f, r_{\text{Sgr A*}} \right) & \simeq & 2 \times 10^{- 13}
%  \left( \frac{T}{7 \times 10^{- 7} \rm{yr}} \right)^{- \frac{1}{4}} \left(
%  \frac{f}{8 \times 10^{- 2} \rm{Hz}} \right)^{- \frac{9}{4}},
%  \hspace{.5em} \text{for} \hspace{.5em} f < \frac{1}{2 \delta t} ~.
%\end{eqnarray}
\begin{eqnarray}
  h_{c, l, \min} \left( f, r_{\text{Sgr*A}} \right) & \simeq & 6.6 \times
  10^{- 15} \left( \frac{T}{1 \rm{yr}} \right)^{- \frac{1}{4}} \left(
  \frac{\delta t}{10 \rm s} \right)^{\frac{1}{4}} \left( \frac{f}{7.6 \times 10^{- 3} \rm{Hz}}
  \right)^{- \frac{1}{2}} \nonumber\\
  &  &  \times \left( \left( \frac{f}{7.6 \times 10^{- 3}
  \rm{Hz}} \right)^{- 3} + \left( \frac{\delta t}{6.3 \times 10^{- 4} \rm s}
  \right) \sigma^2 \right)^{\frac{1}{2}}~,
\end{eqnarray}
where we have let the $N$ be of the order of hundreds, and utilized the results of overlap reduction functions obtained in Sec.~\ref{III}, namely, $\Gamma_{l=2}(\omega,0)\simeq \mathcal{O}(1)$ shown in Fig.~\ref{F7}.  

\begin{figure}
  \includegraphics[width=0.8\linewidth]{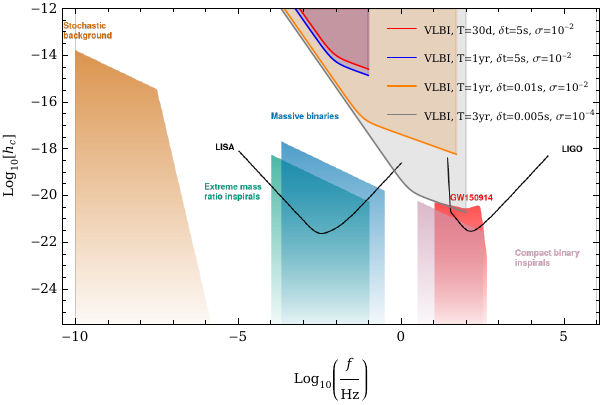}
  \caption{Characteristic strain of gravitational waves in frequency domain for different sources and detectors. The detections of gravitational waves from Sgr A* via VLBI are presented based on our study. Here, the sensitivity curves are given for selected observation duration $T$, cadence $\delta t$ and uncertainty $\sigma$. For comparison, we show the sensitivity curves of LIGO and LISA, and spectral of gravitational wave sources that were given in Ref.~\cite{Moore:2014lga} (also see, \href{http://gwplotter.com/}{gwplotter.com}).
    \label{F11}}
\end{figure}

Fig.~\ref{F11} shows gravitational-wave sensitivity curves of VLBI. The current EHT VLBI could have a high cadence of up to 4 seconds \cite{EventHorizonTelescope:2022apq,EventHorizonTelescope:2022ago}. And with the uncertainty $\sigma\simeq0.01$ for current VLBI \cite{Wielgus:2021peu}, the sensitivity curve is shown in red curve. In this case, the instrumental noise dominates the total noise spectrum $P_{\Delta I}(\omega)$ for $f\gtrsim 0.01$Hz. In LISA frequency band, sensitivity of the EHT VLBI shown in red curve is much inferior to that of LISA. In the LIGO frequency band, the VLBI could be utilized to detect the stochastic gravitational fluctuations, if the cadence is enhanced up to 0.01 seconds \cite{Johnson:2023ynn,10.1117/12.2630313}, and the uncertainty $\sigma$ is reduced to the $10^{-4}$.  The timescales exceeding 1 seconds, the stochastic emission sources would dominate variability of Sgr A* light curves. % The gravitational waves could be observed in the LIGO band  
Theoretically, as the upgrade of the telescopes, $\sigma$ or $\delta t \rightarrow0$, the VLBI-based gravitational fluctuation detection is feasible only on a small scale.

\

\section{Conclusions and discussions}\label{IV}
This paper investigated the two-point correlations of intensity fluctuations on the photon ring, as a result of existence of shock waves in Schwarzschild background.
Following the approaches for gravitational wave detectors \cite{Romano:2016dpx}, we introduced response functions for detecting the stochastic gravitational fluctuations via black hole images, and studied the shape of the overlap reduction functions.
It is found that the shape of the correlations obtained in this paper is distinguishable with that originated from stochastic emission sources.
We also provided an order-of-magnitude estimate for the SNR for detecting the gravitational fluctuations with VLBI. %, in which the stochastic emission sources are considered as noise sources. %A future VLBI with a high cadence of up to 0.01 seconds is promising to detect the gravitational fluctuations in the LIGO frequency band.
%It suggests that the intensity correlations can be used to distinguish between the emission models and black hole geometries in the stochastic regime.  
Theoretically, with the increasing precision of VLBI, the gravitational waves in the LIGO frequency band generated near the supermassive black hole Sgr A* could be detectable in the far future.

This study provided the SNR based on the results of $\Gamma_l(\omega,0)\simeq \mathcal{O}(1)$. Limited by current numerical methods, we did not calculate the $\Gamma_l(\omega,0)$ for all the $\omega$. Simplification or analytical method would be employed in the future studies, in which the sensitivity curves could be reshaped by the overlap reduction functions $\Gamma_l(\omega,0)$.

%The two-point correlations of redshift fluctuations calculated in this paper are the equal-time correlations. Namely, we consider the correlations of two light rays that experience the same time in the Schwarzschild background. To calculate temporal correlations, one should consider critical behavior of light rays near the photon sphere in present of black hole perturbations. It seems not trivial, because the perturbed metric turns to be time-dependent on the photon sphere.

The gravitational fluctuations we considered here are solutions of vacuum Einstein field equations. What kind of physical mechanism can produce the gravitational fluctuations? Besides the shock waves as a results of soft hairs on a black hole, it can also consist of a number of unresolved gravitational wave bursts in astrophysics, such as supernova explosion, or binary mergers occurred in the near regime of a supermassive black hole. Because after these gravitational fluctuations being produced, it would then freely propagate in the vacuum, and can be described by the vacuum Einstein field equations. 
%We presented a different framework from the seed work \cite{Hadar:2020fda} for studying the intensity auto-correlation on the black hole images.
It is worth noting that in Ref.~\cite{Chen:2022kzv}, the framework proposed in Ref.~\cite{Hadar:2020fda} was applied to a black hole surrounded by bosonic clouds. In this case, the oscillating bosonic clouds could also be the physical origin of the shock waves considered in this context. %, they may provide a similar physical picture to our study. %In this sense, it is motivated to present a comprehensive comparison with previous studies \cite{Hadar:2020fda,Chen:2022kzv} in the future.

This paper explored the possibility that VLBI is utilized for gravitational fluctuation detection around a black hole. %For ground-based gravitational wave detectors, such as LIGO, the overlap reduction functions are determined by interferometer design. 
The characteristic signatures described by the overlap reduction functions in Eq.~(\ref{52}) is analogous to that of Hellings-Downs curve for distinguishing gravitational wave signals. Differed from the gravitational wave detectors designed in Minkowski background, in which only the tensor modes exist in Einstein's gravity, the VLBI as gravitational fluctuation detectors can detect both the vector and tensor modes. Namely, as shown in Eq.~(\ref{42}), we have $\mathcal{O}(h_{lm}^{B1,{\rm TT}})\simeq \mathcal{O}(h_{lm}^{B2,{\rm TT}})$  in the near field regions of a black hole. %The observed intensity fluctuations should contain the physical information in the near field of a black hole. 
%Therefore, it is not simply a gravitational wave detectors.
%can encode the gravitational fluctuations in near field region of a black hole. 
%Therefore, the VLBI 

The gravitational fluctuations decay with $r^{-1}$ in the leading order, while the response function of VLBI is proportional to $r^0$. Because the black hole images are composed of light  that carries information of space-time geometry in near regime of a black hole, the decay of the gravitational waves would not suppress the observability of the VLBI observation.  In other words, VLBI observations detect gravitational fluctuations at $r_\text{near photon sphere}$ rather than the $r_\text{distant observers}$. 
This feature suggests that VLBI observations for gravitational fluctuation detection do not merely serve as alternative detectors to current projects for gravitational wave detection. 
%This feature indicates that the VLBI observation for detecting gravitational waves  does not simply serve as an alternative detectors as current projects for gravitational wave detection.
%is not simply an alternative approach to current gravitational wave detections.

\

{\it Acknowledgments. }
The author thanks Prof.~Qing-Guo Huang, Prof. Sai Wang, Prof. Bin Chen, Dr. Minyong Guo, Dr.~Bing Sun, Yehui Hou, and Zhenyu Zhang for useful discussions, and thanks anonymous referees for their valuable comments on the feasibility and observability of the study. This work is supported by the National Nature Science Foundation of China under grant No.~12305073.

\bibliography{ref}

\begin{thebibliography}{50}%
\makeatletter
\providecommand \@ifxundefined [1]{%
 \@ifx{#1\undefined}
}%
\providecommand \@ifnum [1]{%
 \ifnum #1\expandafter \@firstoftwo
 \else \expandafter \@secondoftwo
 \fi
}%
\providecommand \@ifx [1]{%
 \ifx #1\expandafter \@firstoftwo
 \else \expandafter \@secondoftwo
 \fi
}%
\providecommand \natexlab [1]{#1}%
\providecommand \enquote  [1]{``#1''}%
\providecommand \bibnamefont  [1]{#1}%
\providecommand \bibfnamefont [1]{#1}%
\providecommand \citenamefont [1]{#1}%
\providecommand \href@noop [0]{\@secondoftwo}%
\providecommand \href [0]{\begingroup \@sanitize@url \@href}%
\providecommand \@href[1]{\@@startlink{#1}\@@href}%
\providecommand \@@href[1]{\endgroup#1\@@endlink}%
\providecommand \@sanitize@url [0]{\catcode `\\12\catcode `\$12\catcode
  `\&12\catcode `\#12\catcode `\^12\catcode `\_12\catcode `\%12\relax}%
\providecommand \@@startlink[1]{}%
\providecommand \@@endlink[0]{}%
\providecommand \url  [0]{\begingroup\@sanitize@url \@url }%
\providecommand \@url [1]{\endgroup\@href {#1}{\urlprefix }}%
\providecommand \urlprefix  [0]{URL }%
\providecommand \Eprint [0]{\href }%
\providecommand \doibase [0]{http://dx.doi.org/}%
\providecommand \selectlanguage [0]{\@gobble}%
\providecommand \bibinfo  [0]{\@secondoftwo}%
\providecommand \bibfield  [0]{\@secondoftwo}%
\providecommand \translation [1]{[#1]}%
\providecommand \BibitemOpen [0]{}%
\providecommand \bibitemStop [0]{}%
\providecommand \bibitemNoStop [0]{.\EOS\space}%
\providecommand \EOS [0]{\spacefactor3000\relax}%
\providecommand \BibitemShut  [1]{\csname bibitem#1\endcsname}%
\let\auto@bib@innerbib\@empty
%</preamble>
\bibitem [{\citenamefont {Akiyama}\ \emph
  {et~al.}(2019{\natexlab{a}})\citenamefont {Akiyama} \emph
  {et~al.}}]{EventHorizonTelescope:2019dse}%
  \BibitemOpen
  \bibfield  {author} {\bibinfo {author} {\bibfnamefont {K.}~\bibnamefont
  {Akiyama}} \emph {et~al.} (\bibinfo {collaboration} {Event Horizon
  Telescope}),\ }\href {\doibase 10.3847/2041-8213/ab0ec7} {\bibfield
  {journal} {\bibinfo  {journal} {Astrophys. J. Lett.}\ }\textbf {\bibinfo
  {volume} {875}},\ \bibinfo {pages} {L1} (\bibinfo {year}
  {2019}{\natexlab{a}})},\ \Eprint {http://arxiv.org/abs/1906.11238}
  {arXiv:1906.11238 [astro-ph.GA]} \BibitemShut {NoStop}%
\bibitem [{\citenamefont {Akiyama}\ \emph
  {et~al.}(2019{\natexlab{b}})\citenamefont {Akiyama} \emph
  {et~al.}}]{EventHorizonTelescope:2019ggy}%
  \BibitemOpen
  \bibfield  {author} {\bibinfo {author} {\bibfnamefont {K.}~\bibnamefont
  {Akiyama}} \emph {et~al.} (\bibinfo {collaboration} {Event Horizon
  Telescope}),\ }\href {\doibase 10.3847/2041-8213/ab1141} {\bibfield
  {journal} {\bibinfo  {journal} {Astrophys. J. Lett.}\ }\textbf {\bibinfo
  {volume} {875}},\ \bibinfo {pages} {L6} (\bibinfo {year}
  {2019}{\natexlab{b}})},\ \Eprint {http://arxiv.org/abs/1906.11243}
  {arXiv:1906.11243 [astro-ph.GA]} \BibitemShut {NoStop}%
\bibitem [{\citenamefont {Akiyama}\ \emph
  {et~al.}(2019{\natexlab{c}})\citenamefont {Akiyama} \emph
  {et~al.}}]{EventHorizonTelescope:2019pgp}%
  \BibitemOpen
  \bibfield  {author} {\bibinfo {author} {\bibfnamefont {K.}~\bibnamefont
  {Akiyama}} \emph {et~al.} (\bibinfo {collaboration} {Event Horizon
  Telescope}),\ }\href {\doibase 10.3847/2041-8213/ab0f43} {\bibfield
  {journal} {\bibinfo  {journal} {Astrophys. J. Lett.}\ }\textbf {\bibinfo
  {volume} {875}},\ \bibinfo {pages} {L5} (\bibinfo {year}
  {2019}{\natexlab{c}})},\ \Eprint {http://arxiv.org/abs/1906.11242}
  {arXiv:1906.11242 [astro-ph.GA]} \BibitemShut {NoStop}%
\bibitem [{\citenamefont {Kocherlakota}\ \emph {et~al.}(2021)\citenamefont
  {Kocherlakota} \emph {et~al.}}]{EventHorizonTelescope:2021dqv}%
  \BibitemOpen
  \bibfield  {author} {\bibinfo {author} {\bibfnamefont {P.}~\bibnamefont
  {Kocherlakota}} \emph {et~al.} (\bibinfo {collaboration} {Event Horizon
  Telescope}),\ }\href {\doibase 10.1103/PhysRevD.103.104047} {\bibfield
  {journal} {\bibinfo  {journal} {Phys. Rev. D}\ }\textbf {\bibinfo {volume}
  {103}},\ \bibinfo {pages} {104047} (\bibinfo {year} {2021})},\ \Eprint
  {http://arxiv.org/abs/2105.09343} {arXiv:2105.09343 [gr-qc]} \BibitemShut
  {NoStop}%
\bibitem [{\citenamefont {Akiyama}\ \emph
  {et~al.}(2022{\natexlab{a}})\citenamefont {Akiyama} \emph
  {et~al.}}]{EventHorizonTelescope:2022xnr}%
  \BibitemOpen
  \bibfield  {author} {\bibinfo {author} {\bibfnamefont {K.}~\bibnamefont
  {Akiyama}} \emph {et~al.} (\bibinfo {collaboration} {Event Horizon
  Telescope}),\ }\href {\doibase 10.3847/2041-8213/ac6674} {\bibfield
  {journal} {\bibinfo  {journal} {Astrophys. J. Lett.}\ }\textbf {\bibinfo
  {volume} {930}},\ \bibinfo {pages} {L12} (\bibinfo {year}
  {2022}{\natexlab{a}})}\BibitemShut {NoStop}%
\bibitem [{\citenamefont {Johannsen}\ \emph {et~al.}(2016)\citenamefont
  {Johannsen}, \citenamefont {Broderick}, \citenamefont {Plewa}, \citenamefont
  {Chatzopoulos}, \citenamefont {Doeleman}, \citenamefont {Eisenhauer},
  \citenamefont {Fish}, \citenamefont {Genzel}, \citenamefont {Gerhard},\ and\
  \citenamefont {Johnson}}]{Johannsen:2015hib}%
  \BibitemOpen
  \bibfield  {author} {\bibinfo {author} {\bibfnamefont {T.}~\bibnamefont
  {Johannsen}}, \bibinfo {author} {\bibfnamefont {A.~E.}\ \bibnamefont
  {Broderick}}, \bibinfo {author} {\bibfnamefont {P.~M.}\ \bibnamefont
  {Plewa}}, \bibinfo {author} {\bibfnamefont {S.}~\bibnamefont {Chatzopoulos}},
  \bibinfo {author} {\bibfnamefont {S.~S.}\ \bibnamefont {Doeleman}}, \bibinfo
  {author} {\bibfnamefont {F.}~\bibnamefont {Eisenhauer}}, \bibinfo {author}
  {\bibfnamefont {V.~L.}\ \bibnamefont {Fish}}, \bibinfo {author}
  {\bibfnamefont {R.}~\bibnamefont {Genzel}}, \bibinfo {author} {\bibfnamefont
  {O.}~\bibnamefont {Gerhard}}, \ and\ \bibinfo {author} {\bibfnamefont
  {M.~D.}\ \bibnamefont {Johnson}},\ }\href {\doibase
  10.1103/PhysRevLett.116.031101} {\bibfield  {journal} {\bibinfo  {journal}
  {Phys. Rev. Lett.}\ }\textbf {\bibinfo {volume} {116}},\ \bibinfo {pages}
  {031101} (\bibinfo {year} {2016})},\ \Eprint
  {http://arxiv.org/abs/1512.02640} {arXiv:1512.02640 [astro-ph.GA]}
  \BibitemShut {NoStop}%
\bibitem [{\citenamefont {Vagnozzi}\ \emph {et~al.}(2022)\citenamefont
  {Vagnozzi} \emph {et~al.}}]{Vagnozzi:2022moj}%
  \BibitemOpen
  \bibfield  {author} {\bibinfo {author} {\bibfnamefont {S.}~\bibnamefont
  {Vagnozzi}} \emph {et~al.},\ }\href@noop {} {\  (\bibinfo {year} {2022})},\
  \Eprint {http://arxiv.org/abs/2205.07787} {arXiv:2205.07787 [gr-qc]}
  \BibitemShut {NoStop}%
\bibitem [{\citenamefont {Glampedakis}\ and\ \citenamefont
  {Pappas}(2021)}]{Glampedakis:2021oie}%
  \BibitemOpen
  \bibfield  {author} {\bibinfo {author} {\bibfnamefont {K.}~\bibnamefont
  {Glampedakis}}\ and\ \bibinfo {author} {\bibfnamefont {G.}~\bibnamefont
  {Pappas}},\ }\href {\doibase 10.1103/PhysRevD.104.L081503} {\bibfield
  {journal} {\bibinfo  {journal} {Phys. Rev. D}\ }\textbf {\bibinfo {volume}
  {104}},\ \bibinfo {pages} {L081503} (\bibinfo {year} {2021})},\ \Eprint
  {http://arxiv.org/abs/2102.13573} {arXiv:2102.13573 [gr-qc]} \BibitemShut
  {NoStop}%
\bibitem [{\citenamefont {Bardeen}(1973)}]{Bardeen:1973tla}%
  \BibitemOpen
  \bibfield  {author} {\bibinfo {author} {\bibfnamefont {J.~M.}\ \bibnamefont
  {Bardeen}},\ }in\ \href@noop {} {\emph {\bibinfo {booktitle} {{Les Houches
  Summer School of Theoretical Physics}: {Black Holes}}}}\ (\bibinfo {year}
  {1973})\ pp.\ \bibinfo {pages} {215--240}\BibitemShut {NoStop}%
\bibitem [{\citenamefont {Johannsen}\ and\ \citenamefont
  {Psaltis}(2010)}]{Johannsen:2010ru}%
  \BibitemOpen
  \bibfield  {author} {\bibinfo {author} {\bibfnamefont {T.}~\bibnamefont
  {Johannsen}}\ and\ \bibinfo {author} {\bibfnamefont {D.}~\bibnamefont
  {Psaltis}},\ }\href {\doibase 10.1088/0004-637X/718/1/446} {\bibfield
  {journal} {\bibinfo  {journal} {Astrophys. J.}\ }\textbf {\bibinfo {volume}
  {718}},\ \bibinfo {pages} {446} (\bibinfo {year} {2010})},\ \Eprint
  {http://arxiv.org/abs/1005.1931} {arXiv:1005.1931 [astro-ph.HE]} \BibitemShut
  {NoStop}%
\bibitem [{\citenamefont {Cunha}\ \emph {et~al.}(2015)\citenamefont {Cunha},
  \citenamefont {Herdeiro}, \citenamefont {Radu},\ and\ \citenamefont
  {Runarsson}}]{Cunha:2015yba}%
  \BibitemOpen
  \bibfield  {author} {\bibinfo {author} {\bibfnamefont {P.~V.~P.}\
  \bibnamefont {Cunha}}, \bibinfo {author} {\bibfnamefont {C.~A.~R.}\
  \bibnamefont {Herdeiro}}, \bibinfo {author} {\bibfnamefont {E.}~\bibnamefont
  {Radu}}, \ and\ \bibinfo {author} {\bibfnamefont {H.~F.}\ \bibnamefont
  {Runarsson}},\ }\href {\doibase 10.1103/PhysRevLett.115.211102} {\bibfield
  {journal} {\bibinfo  {journal} {Phys. Rev. Lett.}\ }\textbf {\bibinfo
  {volume} {115}},\ \bibinfo {pages} {211102} (\bibinfo {year} {2015})},\
  \Eprint {http://arxiv.org/abs/1509.00021} {arXiv:1509.00021 [gr-qc]}
  \BibitemShut {NoStop}%
\bibitem [{\citenamefont {Gan}\ \emph {et~al.}(2021)\citenamefont {Gan},
  \citenamefont {Wang}, \citenamefont {Wu},\ and\ \citenamefont
  {Yang}}]{Gan:2021xdl}%
  \BibitemOpen
  \bibfield  {author} {\bibinfo {author} {\bibfnamefont {Q.}~\bibnamefont
  {Gan}}, \bibinfo {author} {\bibfnamefont {P.}~\bibnamefont {Wang}}, \bibinfo
  {author} {\bibfnamefont {H.}~\bibnamefont {Wu}}, \ and\ \bibinfo {author}
  {\bibfnamefont {H.}~\bibnamefont {Yang}},\ }\href {\doibase
  10.1103/PhysRevD.104.044049} {\bibfield  {journal} {\bibinfo  {journal}
  {Phys. Rev. D}\ }\textbf {\bibinfo {volume} {104}},\ \bibinfo {pages}
  {044049} (\bibinfo {year} {2021})},\ \Eprint
  {http://arxiv.org/abs/2105.11770} {arXiv:2105.11770 [gr-qc]} \BibitemShut
  {NoStop}%
\bibitem [{\citenamefont {Ghosh}\ and\ \citenamefont
  {Bhattacharyya}(2022)}]{Ghosh:2022mka}%
  \BibitemOpen
  \bibfield  {author} {\bibinfo {author} {\bibfnamefont {S.}~\bibnamefont
  {Ghosh}}\ and\ \bibinfo {author} {\bibfnamefont {A.}~\bibnamefont
  {Bhattacharyya}},\ }\href {\doibase 10.1088/1475-7516/2022/11/006} {\bibfield
   {journal} {\bibinfo  {journal} {JCAP}\ }\textbf {\bibinfo {volume} {11}},\
  \bibinfo {pages} {006} (\bibinfo {year} {2022})},\ \Eprint
  {http://arxiv.org/abs/2206.09954} {arXiv:2206.09954 [gr-qc]} \BibitemShut
  {NoStop}%
\bibitem [{\citenamefont {Rosa}\ and\ \citenamefont
  {Rubiera-Garcia}(2022)}]{Rosa:2022tfv}%
  \BibitemOpen
  \bibfield  {author} {\bibinfo {author} {\bibfnamefont {J.~a.~L.}\
  \bibnamefont {Rosa}}\ and\ \bibinfo {author} {\bibfnamefont {D.}~\bibnamefont
  {Rubiera-Garcia}},\ }\href {\doibase 10.1103/PhysRevD.106.084004} {\bibfield
  {journal} {\bibinfo  {journal} {Phys. Rev. D}\ }\textbf {\bibinfo {volume}
  {106}},\ \bibinfo {pages} {084004} (\bibinfo {year} {2022})},\ \Eprint
  {http://arxiv.org/abs/2204.12949} {arXiv:2204.12949 [gr-qc]} \BibitemShut
  {NoStop}%
\bibitem [{\citenamefont {Chowdhuri}\ and\ \citenamefont
  {Bhattacharyya}(2021)}]{Chowdhuri:2020ipb}%
  \BibitemOpen
  \bibfield  {author} {\bibinfo {author} {\bibfnamefont {A.}~\bibnamefont
  {Chowdhuri}}\ and\ \bibinfo {author} {\bibfnamefont {A.}~\bibnamefont
  {Bhattacharyya}},\ }\href {\doibase 10.1103/PhysRevD.104.064039} {\bibfield
  {journal} {\bibinfo  {journal} {Phys. Rev. D}\ }\textbf {\bibinfo {volume}
  {104}},\ \bibinfo {pages} {064039} (\bibinfo {year} {2021})},\ \Eprint
  {http://arxiv.org/abs/2012.12914} {arXiv:2012.12914 [gr-qc]} \BibitemShut
  {NoStop}%
\bibitem [{\citenamefont {Perlick}\ and\ \citenamefont
  {Tsupko}(2022)}]{Perlick:2021aok}%
  \BibitemOpen
  \bibfield  {author} {\bibinfo {author} {\bibfnamefont {V.}~\bibnamefont
  {Perlick}}\ and\ \bibinfo {author} {\bibfnamefont {O.~Y.}\ \bibnamefont
  {Tsupko}},\ }\href {\doibase 10.1016/j.physrep.2021.10.004} {\bibfield
  {journal} {\bibinfo  {journal} {Phys. Rept.}\ }\textbf {\bibinfo {volume}
  {947}},\ \bibinfo {pages} {1} (\bibinfo {year} {2022})},\ \Eprint
  {http://arxiv.org/abs/2105.07101} {arXiv:2105.07101 [gr-qc]} \BibitemShut
  {NoStop}%
\bibitem [{\citenamefont {Chang}\ and\ \citenamefont
  {Zhu}(2020)}]{Chang:2020miq}%
  \BibitemOpen
  \bibfield  {author} {\bibinfo {author} {\bibfnamefont {Z.}~\bibnamefont
  {Chang}}\ and\ \bibinfo {author} {\bibfnamefont {Q.-H.}\ \bibnamefont
  {Zhu}},\ }\href {\doibase 10.1103/PhysRevD.101.084029} {\bibfield  {journal}
  {\bibinfo  {journal} {Phys. Rev. D}\ }\textbf {\bibinfo {volume} {101}},\
  \bibinfo {pages} {084029} (\bibinfo {year} {2020})},\ \Eprint
  {http://arxiv.org/abs/2001.05175} {arXiv:2001.05175 [gr-qc]} \BibitemShut
  {NoStop}%
\bibitem [{\citenamefont {Gralla}\ \emph {et~al.}(2019)\citenamefont {Gralla},
  \citenamefont {Holz},\ and\ \citenamefont {Wald}}]{Gralla:2019xty}%
  \BibitemOpen
  \bibfield  {author} {\bibinfo {author} {\bibfnamefont {S.~E.}\ \bibnamefont
  {Gralla}}, \bibinfo {author} {\bibfnamefont {D.~E.}\ \bibnamefont {Holz}}, \
  and\ \bibinfo {author} {\bibfnamefont {R.~M.}\ \bibnamefont {Wald}},\ }\href
  {\doibase 10.1103/PhysRevD.100.024018} {\bibfield  {journal} {\bibinfo
  {journal} {Phys. Rev. D}\ }\textbf {\bibinfo {volume} {100}},\ \bibinfo
  {pages} {024018} (\bibinfo {year} {2019})},\ \Eprint
  {http://arxiv.org/abs/1906.00873} {arXiv:1906.00873 [astro-ph.HE]}
  \BibitemShut {NoStop}%
\bibitem [{\citenamefont {Narayan}\ \emph {et~al.}(2019)\citenamefont
  {Narayan}, \citenamefont {Johnson},\ and\ \citenamefont
  {Gammie}}]{Narayan:2019imo}%
  \BibitemOpen
  \bibfield  {author} {\bibinfo {author} {\bibfnamefont {R.}~\bibnamefont
  {Narayan}}, \bibinfo {author} {\bibfnamefont {M.~D.}\ \bibnamefont
  {Johnson}}, \ and\ \bibinfo {author} {\bibfnamefont {C.~F.}\ \bibnamefont
  {Gammie}},\ }\href {\doibase 10.3847/2041-8213/ab518c} {\bibfield  {journal}
  {\bibinfo  {journal} {Astrophys. J. Lett.}\ }\textbf {\bibinfo {volume}
  {885}},\ \bibinfo {pages} {L33} (\bibinfo {year} {2019})},\ \Eprint
  {http://arxiv.org/abs/1910.02957} {arXiv:1910.02957 [astro-ph.HE]}
  \BibitemShut {NoStop}%
\bibitem [{\citenamefont {Bronzwaer}\ and\ \citenamefont
  {Falcke}(2021)}]{Bronzwaer:2021lzo}%
  \BibitemOpen
  \bibfield  {author} {\bibinfo {author} {\bibfnamefont {T.}~\bibnamefont
  {Bronzwaer}}\ and\ \bibinfo {author} {\bibfnamefont {H.}~\bibnamefont
  {Falcke}},\ }\href {\doibase 10.3847/1538-4357/ac1738} {\bibfield  {journal}
  {\bibinfo  {journal} {Astrophys. J.}\ }\textbf {\bibinfo {volume} {920}},\
  \bibinfo {pages} {155} (\bibinfo {year} {2021})},\ \Eprint
  {http://arxiv.org/abs/2108.03966} {arXiv:2108.03966 [astro-ph.HE]}
  \BibitemShut {NoStop}%
\bibitem [{\citenamefont {Kocherlakota}\ and\ \citenamefont
  {Rezzolla}(2022)}]{Kocherlakota:2022jnz}%
  \BibitemOpen
  \bibfield  {author} {\bibinfo {author} {\bibfnamefont {P.}~\bibnamefont
  {Kocherlakota}}\ and\ \bibinfo {author} {\bibfnamefont {L.}~\bibnamefont
  {Rezzolla}},\ }\href {\doibase 10.1093/mnras/stac891} {\bibfield  {journal}
  {\bibinfo  {journal} {Mon. Not. Roy. Astron. Soc.}\ }\textbf {\bibinfo
  {volume} {513}},\ \bibinfo {pages} {1229} (\bibinfo {year} {2022})},\ \Eprint
  {http://arxiv.org/abs/2201.05641} {arXiv:2201.05641 [gr-qc]} \BibitemShut
  {NoStop}%
\bibitem [{\citenamefont {Gralla}\ \emph {et~al.}(2020)\citenamefont {Gralla},
  \citenamefont {Lupsasca},\ and\ \citenamefont {Marrone}}]{Gralla:2020srx}%
  \BibitemOpen
  \bibfield  {author} {\bibinfo {author} {\bibfnamefont {S.~E.}\ \bibnamefont
  {Gralla}}, \bibinfo {author} {\bibfnamefont {A.}~\bibnamefont {Lupsasca}}, \
  and\ \bibinfo {author} {\bibfnamefont {D.~P.}\ \bibnamefont {Marrone}},\
  }\href {\doibase 10.1103/PhysRevD.102.124004} {\bibfield  {journal} {\bibinfo
   {journal} {Phys. Rev. D}\ }\textbf {\bibinfo {volume} {102}},\ \bibinfo
  {pages} {124004} (\bibinfo {year} {2020})},\ \Eprint
  {http://arxiv.org/abs/2008.03879} {arXiv:2008.03879 [gr-qc]} \BibitemShut
  {NoStop}%
\bibitem [{\citenamefont {Gralla}\ and\ \citenamefont
  {Lupsasca}(2020)}]{Gralla:2020yvo}%
  \BibitemOpen
  \bibfield  {author} {\bibinfo {author} {\bibfnamefont {S.~E.}\ \bibnamefont
  {Gralla}}\ and\ \bibinfo {author} {\bibfnamefont {A.}~\bibnamefont
  {Lupsasca}},\ }\href {\doibase 10.1103/PhysRevD.102.124003} {\bibfield
  {journal} {\bibinfo  {journal} {Phys. Rev. D}\ }\textbf {\bibinfo {volume}
  {102}},\ \bibinfo {pages} {124003} (\bibinfo {year} {2020})},\ \Eprint
  {http://arxiv.org/abs/2007.10336} {arXiv:2007.10336 [gr-qc]} \BibitemShut
  {NoStop}%
\bibitem [{\citenamefont {Johnson}\ \emph {et~al.}(2020)\citenamefont {Johnson}
  \emph {et~al.}}]{Johnson:2019ljv}%
  \BibitemOpen
  \bibfield  {author} {\bibinfo {author} {\bibfnamefont {M.~D.}\ \bibnamefont
  {Johnson}} \emph {et~al.},\ }\href {\doibase 10.1126/sciadv.aaz1310}
  {\bibfield  {journal} {\bibinfo  {journal} {Sci. Adv.}\ }\textbf {\bibinfo
  {volume} {6}},\ \bibinfo {pages} {eaaz1310} (\bibinfo {year} {2020})},\
  \Eprint {http://arxiv.org/abs/1907.04329} {arXiv:1907.04329 [astro-ph.IM]}
  \BibitemShut {NoStop}%
\bibitem [{\citenamefont {Broderick}\ \emph {et~al.}(2022)\citenamefont
  {Broderick}, \citenamefont {Tiede}, \citenamefont {Pesce},\ and\
  \citenamefont {Gold}}]{Broderick:2021ohx}%
  \BibitemOpen
  \bibfield  {author} {\bibinfo {author} {\bibfnamefont {A.~E.}\ \bibnamefont
  {Broderick}}, \bibinfo {author} {\bibfnamefont {P.}~\bibnamefont {Tiede}},
  \bibinfo {author} {\bibfnamefont {D.~W.}\ \bibnamefont {Pesce}}, \ and\
  \bibinfo {author} {\bibfnamefont {R.}~\bibnamefont {Gold}},\ }\href {\doibase
  10.3847/1538-4357/ac4970} {\bibfield  {journal} {\bibinfo  {journal}
  {Astrophys. J.}\ }\textbf {\bibinfo {volume} {927}},\ \bibinfo {pages} {6}
  (\bibinfo {year} {2022})},\ \Eprint {http://arxiv.org/abs/2105.09962}
  {arXiv:2105.09962 [astro-ph.HE]} \BibitemShut {NoStop}%
\bibitem [{\citenamefont {Wielgus}(2021)}]{Wielgus:2021peu}%
  \BibitemOpen
  \bibfield  {author} {\bibinfo {author} {\bibfnamefont {M.}~\bibnamefont
  {Wielgus}},\ }\href {\doibase 10.1103/PhysRevD.104.124058} {\bibfield
  {journal} {\bibinfo  {journal} {Phys. Rev. D}\ }\textbf {\bibinfo {volume}
  {104}},\ \bibinfo {pages} {124058} (\bibinfo {year} {2021})},\ \Eprint
  {http://arxiv.org/abs/2109.10840} {arXiv:2109.10840 [gr-qc]} \BibitemShut
  {NoStop}%
\bibitem [{\citenamefont {Tsupko}(2022)}]{Tsupko:2022kwi}%
  \BibitemOpen
  \bibfield  {author} {\bibinfo {author} {\bibfnamefont {O.~Y.}\ \bibnamefont
  {Tsupko}},\ }\href {\doibase 10.1103/PhysRevD.106.064033} {\bibfield
  {journal} {\bibinfo  {journal} {Phys. Rev. D}\ }\textbf {\bibinfo {volume}
  {106}},\ \bibinfo {pages} {064033} (\bibinfo {year} {2022})},\ \Eprint
  {http://arxiv.org/abs/2208.02084} {arXiv:2208.02084 [gr-qc]} \BibitemShut
  {NoStop}%
\bibitem [{\citenamefont {Gralla}(2021)}]{Gralla:2020pra}%
  \BibitemOpen
  \bibfield  {author} {\bibinfo {author} {\bibfnamefont {S.~E.}\ \bibnamefont
  {Gralla}},\ }\href {\doibase 10.1103/PhysRevD.103.024023} {\bibfield
  {journal} {\bibinfo  {journal} {Phys. Rev. D}\ }\textbf {\bibinfo {volume}
  {103}},\ \bibinfo {pages} {024023} (\bibinfo {year} {2021})},\ \Eprint
  {http://arxiv.org/abs/2010.08557} {arXiv:2010.08557 [astro-ph.HE]}
  \BibitemShut {NoStop}%
\bibitem [{\citenamefont {Vincent}\ \emph {et~al.}(2021)\citenamefont
  {Vincent}, \citenamefont {Wielgus}, \citenamefont {Abramowicz}, \citenamefont
  {Gourgoulhon}, \citenamefont {Lasota}, \citenamefont {Paumard},\ and\
  \citenamefont {Perrin}}]{Vincent:2020dij}%
  \BibitemOpen
  \bibfield  {author} {\bibinfo {author} {\bibfnamefont {F.~H.}\ \bibnamefont
  {Vincent}}, \bibinfo {author} {\bibfnamefont {M.}~\bibnamefont {Wielgus}},
  \bibinfo {author} {\bibfnamefont {M.~A.}\ \bibnamefont {Abramowicz}},
  \bibinfo {author} {\bibfnamefont {E.}~\bibnamefont {Gourgoulhon}}, \bibinfo
  {author} {\bibfnamefont {J.~P.}\ \bibnamefont {Lasota}}, \bibinfo {author}
  {\bibfnamefont {T.}~\bibnamefont {Paumard}}, \ and\ \bibinfo {author}
  {\bibfnamefont {G.}~\bibnamefont {Perrin}},\ }\href {\doibase
  10.1051/0004-6361/202037787} {\bibfield  {journal} {\bibinfo  {journal}
  {Astron. Astrophys.}\ }\textbf {\bibinfo {volume} {646}},\ \bibinfo {pages}
  {A37} (\bibinfo {year} {2021})},\ \Eprint {http://arxiv.org/abs/2002.09226}
  {arXiv:2002.09226 [gr-qc]} \BibitemShut {NoStop}%
\bibitem [{\citenamefont {Tiede}\ \emph {et~al.}(2022)\citenamefont {Tiede},
  \citenamefont {Johnson}, \citenamefont {Pesce}, \citenamefont {Palumbo},
  \citenamefont {Chang},\ and\ \citenamefont {Galison}}]{Tiede:2022grp}%
  \BibitemOpen
  \bibfield  {author} {\bibinfo {author} {\bibfnamefont {P.}~\bibnamefont
  {Tiede}}, \bibinfo {author} {\bibfnamefont {M.~D.}\ \bibnamefont {Johnson}},
  \bibinfo {author} {\bibfnamefont {D.~W.}\ \bibnamefont {Pesce}}, \bibinfo
  {author} {\bibfnamefont {D.~C.~M.}\ \bibnamefont {Palumbo}}, \bibinfo
  {author} {\bibfnamefont {D.~O.}\ \bibnamefont {Chang}}, \ and\ \bibinfo
  {author} {\bibfnamefont {P.}~\bibnamefont {Galison}},\ }\href@noop {} {\
  (\bibinfo {year} {2022})},\ \Eprint {http://arxiv.org/abs/2210.13498}
  {arXiv:2210.13498 [astro-ph.HE]} \BibitemShut {NoStop}%
\bibitem [{\citenamefont {Hadar}\ \emph {et~al.}(2021)\citenamefont {Hadar},
  \citenamefont {Johnson}, \citenamefont {Lupsasca},\ and\ \citenamefont
  {Wong}}]{Hadar:2020fda}%
  \BibitemOpen
  \bibfield  {author} {\bibinfo {author} {\bibfnamefont {S.}~\bibnamefont
  {Hadar}}, \bibinfo {author} {\bibfnamefont {M.~D.}\ \bibnamefont {Johnson}},
  \bibinfo {author} {\bibfnamefont {A.}~\bibnamefont {Lupsasca}}, \ and\
  \bibinfo {author} {\bibfnamefont {G.~N.}\ \bibnamefont {Wong}},\ }\href
  {\doibase 10.1103/PhysRevD.103.104038} {\bibfield  {journal} {\bibinfo
  {journal} {Phys. Rev. D}\ }\textbf {\bibinfo {volume} {103}},\ \bibinfo
  {pages} {104038} (\bibinfo {year} {2021})},\ \Eprint
  {http://arxiv.org/abs/2010.03683} {arXiv:2010.03683 [gr-qc]} \BibitemShut
  {NoStop}%
\bibitem [{\citenamefont {Chen}\ \emph {et~al.}(2023)\citenamefont {Chen},
  \citenamefont {Xue}, \citenamefont {Brito},\ and\ \citenamefont
  {Cardoso}}]{Chen:2022kzv}%
  \BibitemOpen
  \bibfield  {author} {\bibinfo {author} {\bibfnamefont {Y.}~\bibnamefont
  {Chen}}, \bibinfo {author} {\bibfnamefont {X.}~\bibnamefont {Xue}}, \bibinfo
  {author} {\bibfnamefont {R.}~\bibnamefont {Brito}}, \ and\ \bibinfo {author}
  {\bibfnamefont {V.}~\bibnamefont {Cardoso}},\ }\href {\doibase
  10.1103/PhysRevLett.130.111401} {\bibfield  {journal} {\bibinfo  {journal}
  {Phys. Rev. Lett.}\ }\textbf {\bibinfo {volume} {130}},\ \bibinfo {pages}
  {111401} (\bibinfo {year} {2023})},\ \Eprint
  {http://arxiv.org/abs/2211.03794} {arXiv:2211.03794 [gr-qc]} \BibitemShut
  {NoStop}%
\bibitem [{\citenamefont {Sarkar}\ \emph {et~al.}(2022)\citenamefont {Sarkar},
  \citenamefont {Kumar},\ and\ \citenamefont {Bhattacharjee}}]{Sarkar:2021djs}%
  \BibitemOpen
  \bibfield  {author} {\bibinfo {author} {\bibfnamefont {S.}~\bibnamefont
  {Sarkar}}, \bibinfo {author} {\bibfnamefont {S.}~\bibnamefont {Kumar}}, \
  and\ \bibinfo {author} {\bibfnamefont {S.}~\bibnamefont {Bhattacharjee}},\
  }\href {\doibase 10.1103/PhysRevD.105.084001} {\bibfield  {journal} {\bibinfo
   {journal} {Phys. Rev. D}\ }\textbf {\bibinfo {volume} {105}},\ \bibinfo
  {pages} {084001} (\bibinfo {year} {2022})},\ \Eprint
  {http://arxiv.org/abs/2110.03547} {arXiv:2110.03547 [gr-qc]} \BibitemShut
  {NoStop}%
\bibitem [{\citenamefont {Lin}\ \emph {et~al.}(2022)\citenamefont {Lin},
  \citenamefont {Patel},\ and\ \citenamefont {Pu}}]{Lin:2022ksb}%
  \BibitemOpen
  \bibfield  {author} {\bibinfo {author} {\bibfnamefont {F.-L.}\ \bibnamefont
  {Lin}}, \bibinfo {author} {\bibfnamefont {A.}~\bibnamefont {Patel}}, \ and\
  \bibinfo {author} {\bibfnamefont {H.-Y.}\ \bibnamefont {Pu}},\ }\href@noop {}
  {\  (\bibinfo {year} {2022})},\ \Eprint {http://arxiv.org/abs/2202.13559}
  {arXiv:2202.13559 [gr-qc]} \BibitemShut {NoStop}%
\bibitem [{\citenamefont {Zhu}\ \emph {et~al.}(2022)\citenamefont {Zhu},
  \citenamefont {Han},\ and\ \citenamefont {Huang}}]{Zhu:2022shb}%
  \BibitemOpen
  \bibfield  {author} {\bibinfo {author} {\bibfnamefont {Q.-H.}\ \bibnamefont
  {Zhu}}, \bibinfo {author} {\bibfnamefont {Y.-X.}\ \bibnamefont {Han}}, \ and\
  \bibinfo {author} {\bibfnamefont {Q.-G.}\ \bibnamefont {Huang}},\ }\href@noop
  {} {\  (\bibinfo {year} {2022})},\ \Eprint {http://arxiv.org/abs/2205.14554}
  {arXiv:2205.14554 [gr-qc]} \BibitemShut {NoStop}%
\bibitem [{\citenamefont {Hawking}\ \emph {et~al.}(2016)\citenamefont
  {Hawking}, \citenamefont {Perry},\ and\ \citenamefont
  {Strominger}}]{Hawking:2016msc}%
  \BibitemOpen
  \bibfield  {author} {\bibinfo {author} {\bibfnamefont {S.~W.}\ \bibnamefont
  {Hawking}}, \bibinfo {author} {\bibfnamefont {M.~J.}\ \bibnamefont {Perry}},
  \ and\ \bibinfo {author} {\bibfnamefont {A.}~\bibnamefont {Strominger}},\
  }\href {\doibase 10.1103/PhysRevLett.116.231301} {\bibfield  {journal}
  {\bibinfo  {journal} {Phys. Rev. Lett.}\ }\textbf {\bibinfo {volume} {116}},\
  \bibinfo {pages} {231301} (\bibinfo {year} {2016})},\ \Eprint
  {http://arxiv.org/abs/1601.00921} {arXiv:1601.00921 [hep-th]} \BibitemShut
  {NoStop}%
\bibitem [{\citenamefont {Strominger}\ and\ \citenamefont
  {Zhiboedov}(2016)}]{Strominger:2014pwa}%
  \BibitemOpen
  \bibfield  {author} {\bibinfo {author} {\bibfnamefont {A.}~\bibnamefont
  {Strominger}}\ and\ \bibinfo {author} {\bibfnamefont {A.}~\bibnamefont
  {Zhiboedov}},\ }\href {\doibase 10.1007/JHEP01(2016)086} {\bibfield
  {journal} {\bibinfo  {journal} {JHEP}\ }\textbf {\bibinfo {volume} {01}},\
  \bibinfo {pages} {086} (\bibinfo {year} {2016})},\ \Eprint
  {http://arxiv.org/abs/1411.5745} {arXiv:1411.5745 [hep-th]} \BibitemShut
  {NoStop}%
\bibitem [{\citenamefont {Comp\`ere}\ and\ \citenamefont
  {Long}(2016)}]{Compere:2016hzt}%
  \BibitemOpen
  \bibfield  {author} {\bibinfo {author} {\bibfnamefont {G.}~\bibnamefont
  {Comp\`ere}}\ and\ \bibinfo {author} {\bibfnamefont {J.}~\bibnamefont
  {Long}},\ }\href {\doibase 10.1088/0264-9381/33/19/195001} {\bibfield
  {journal} {\bibinfo  {journal} {Class. Quant. Grav.}\ }\textbf {\bibinfo
  {volume} {33}},\ \bibinfo {pages} {195001} (\bibinfo {year} {2016})},\
  \Eprint {http://arxiv.org/abs/1602.05197} {arXiv:1602.05197 [gr-qc]}
  \BibitemShut {NoStop}%
\bibitem [{\citenamefont {Hawking}\ \emph {et~al.}(2017)\citenamefont
  {Hawking}, \citenamefont {Perry},\ and\ \citenamefont
  {Strominger}}]{Hawking:2016sgy}%
  \BibitemOpen
  \bibfield  {author} {\bibinfo {author} {\bibfnamefont {S.~W.}\ \bibnamefont
  {Hawking}}, \bibinfo {author} {\bibfnamefont {M.~J.}\ \bibnamefont {Perry}},
  \ and\ \bibinfo {author} {\bibfnamefont {A.}~\bibnamefont {Strominger}},\
  }\href {\doibase 10.1007/JHEP05(2017)161} {\bibfield  {journal} {\bibinfo
  {journal} {JHEP}\ }\textbf {\bibinfo {volume} {05}},\ \bibinfo {pages} {161}
  (\bibinfo {year} {2017})},\ \Eprint {http://arxiv.org/abs/1611.09175}
  {arXiv:1611.09175 [hep-th]} \BibitemShut {NoStop}%
\bibitem [{\citenamefont {Fiziev}(2006)}]{Fiziev:2005ki}%
  \BibitemOpen
  \bibfield  {author} {\bibinfo {author} {\bibfnamefont {P.~P.}\ \bibnamefont
  {Fiziev}},\ }\href {\doibase 10.1088/0264-9381/23/7/015} {\bibfield
  {journal} {\bibinfo  {journal} {Class. Quant. Grav.}\ }\textbf {\bibinfo
  {volume} {23}},\ \bibinfo {pages} {2447} (\bibinfo {year} {2006})},\ \Eprint
  {http://arxiv.org/abs/gr-qc/0509123} {arXiv:gr-qc/0509123} \BibitemShut
  {NoStop}%
\bibitem [{\citenamefont {Romano}\ and\ \citenamefont
  {Cornish}(2017)}]{Romano:2016dpx}%
  \BibitemOpen
  \bibfield  {author} {\bibinfo {author} {\bibfnamefont {J.~D.}\ \bibnamefont
  {Romano}}\ and\ \bibinfo {author} {\bibfnamefont {N.~J.}\ \bibnamefont
  {Cornish}},\ }\href {\doibase 10.1007/s41114-017-0004-1} {\bibfield
  {journal} {\bibinfo  {journal} {Living Rev. Rel.}\ }\textbf {\bibinfo
  {volume} {20}},\ \bibinfo {pages} {2} (\bibinfo {year} {2017})},\ \Eprint
  {http://arxiv.org/abs/1608.06889} {arXiv:1608.06889 [gr-qc]} \BibitemShut
  {NoStop}%
\bibitem [{\citenamefont {Maggiore}(2018)}]{Maggiore:2018sht}%
  \BibitemOpen
  \bibfield  {author} {\bibinfo {author} {\bibfnamefont {M.}~\bibnamefont
  {Maggiore}},\ }\href@noop {} {\emph {\bibinfo {title} {{Gravitational Waves.
  Vol. 2: Astrophysics and Cosmology}}}}\ (\bibinfo  {publisher} {Oxford
  University Press},\ \bibinfo {year} {2018})\BibitemShut {NoStop}%
\bibitem [{\citenamefont {Hellings}\ and\ \citenamefont
  {Downs}(1983)}]{Hellings:1983fr}%
  \BibitemOpen
  \bibfield  {author} {\bibinfo {author} {\bibfnamefont {R.~w.}\ \bibnamefont
  {Hellings}}\ and\ \bibinfo {author} {\bibfnamefont {G.~s.}\ \bibnamefont
  {Downs}},\ }\href {\doibase 10.1086/183954} {\bibfield  {journal} {\bibinfo
  {journal} {Astrophys. J. Lett.}\ }\textbf {\bibinfo {volume} {265}},\
  \bibinfo {pages} {L39} (\bibinfo {year} {1983})}\BibitemShut {NoStop}%
\bibitem [{\citenamefont {McDonough}\ and\ \citenamefont
  {Whalen}(1995)}]{10.5555/559989}%
  \BibitemOpen
  \bibfield  {author} {\bibinfo {author} {\bibfnamefont {R.~N.}\ \bibnamefont
  {McDonough}}\ and\ \bibinfo {author} {\bibfnamefont {A.~D.}\ \bibnamefont
  {Whalen}},\ }\href@noop {} {\emph {\bibinfo {title} {Detection of Signals in
  Noise}}},\ \bibinfo {edition} {2nd}\ ed.\ (\bibinfo  {publisher} {Academic
  Press, Inc.},\ \bibinfo {address} {USA},\ \bibinfo {year} {1995})\BibitemShut
  {NoStop}%
\bibitem [{\citenamefont {Wielgus}\ \emph {et~al.}(2022)\citenamefont {Wielgus}
  \emph {et~al.}}]{EventHorizonTelescope:2022ago}%
  \BibitemOpen
  \bibfield  {author} {\bibinfo {author} {\bibfnamefont {M.}~\bibnamefont
  {Wielgus}} \emph {et~al.} (\bibinfo {collaboration} {Event Horizon
  Telescope}),\ }\href {\doibase 10.3847/2041-8213/ac6428} {\bibfield
  {journal} {\bibinfo  {journal} {Astrophys. J. Lett.}\ }\textbf {\bibinfo
  {volume} {930}},\ \bibinfo {pages} {L19} (\bibinfo {year} {2022})},\ \Eprint
  {http://arxiv.org/abs/2207.06829} {arXiv:2207.06829 [astro-ph.HE]}
  \BibitemShut {NoStop}%
\bibitem [{\citenamefont {Georgiev}\ \emph {et~al.}(2022)\citenamefont
  {Georgiev} \emph {et~al.}}]{EventHorizonTelescope:2022okn}%
  \BibitemOpen
  \bibfield  {author} {\bibinfo {author} {\bibfnamefont {B.}~\bibnamefont
  {Georgiev}} \emph {et~al.} (\bibinfo {collaboration} {Event Horizon
  Telescope}),\ }\href {\doibase 10.3847/2041-8213/ac65eb} {\bibfield
  {journal} {\bibinfo  {journal} {Astrophys. J. Lett.}\ }\textbf {\bibinfo
  {volume} {930}},\ \bibinfo {pages} {L20} (\bibinfo {year}
  {2022})}\BibitemShut {NoStop}%
\bibitem [{\citenamefont {Akiyama}\ \emph
  {et~al.}(2022{\natexlab{b}})\citenamefont {Akiyama} \emph
  {et~al.}}]{EventHorizonTelescope:2022apq}%
  \BibitemOpen
  \bibfield  {author} {\bibinfo {author} {\bibfnamefont {K.}~\bibnamefont
  {Akiyama}} \emph {et~al.} (\bibinfo {collaboration} {Event Horizon
  Telescope}),\ }\href {\doibase 10.3847/2041-8213/ac6675} {\bibfield
  {journal} {\bibinfo  {journal} {Astrophys. J. Lett.}\ }\textbf {\bibinfo
  {volume} {930}},\ \bibinfo {pages} {L13} (\bibinfo {year}
  {2022}{\natexlab{b}})}\BibitemShut {NoStop}%
\bibitem [{\citenamefont {Johnson}\ \emph {et~al.}(2023)\citenamefont {Johnson}
  \emph {et~al.}}]{Johnson:2023ynn}%
  \BibitemOpen
  \bibfield  {author} {\bibinfo {author} {\bibfnamefont {M.~D.}\ \bibnamefont
  {Johnson}} \emph {et~al.},\ }\href {\doibase 10.3390/galaxies11030061}
  {\bibfield  {journal} {\bibinfo  {journal} {Galaxies}\ }\textbf {\bibinfo
  {volume} {11}},\ \bibinfo {pages} {61} (\bibinfo {year} {2023})},\ \Eprint
  {http://arxiv.org/abs/2304.11188} {arXiv:2304.11188 [astro-ph.HE]}
  \BibitemShut {NoStop}%
\bibitem [{\citenamefont {Kurczynski}\ \emph {et~al.}(2022)\citenamefont
  {Kurczynski}, \citenamefont {Johnson}, \citenamefont {Doeleman},
  \citenamefont {Haworth}, \citenamefont {Peretz}, \citenamefont {Sridharan},
  \citenamefont {Bilyeu}, \citenamefont {Blackburn}, \citenamefont {Boroson},
  \citenamefont {Brosius}, \citenamefont {Butler}, \citenamefont {Caplan},
  \citenamefont {Chatterjee}, \citenamefont {Cheimets}, \citenamefont
  {D'Orazio}, \citenamefont {Essinger-Hileman}, \citenamefont {Galison},
  \citenamefont {Gamble}, \citenamefont {Hadar}, \citenamefont {Hoerbelt},
  \citenamefont {Jiao}, \citenamefont {Kauffmann}, \citenamefont {Lafon},
  \citenamefont {Ma}, \citenamefont {Melnick}, \citenamefont {Newbury},
  \citenamefont {Noble}, \citenamefont {Palumbo}, \citenamefont {Paritsky},
  \citenamefont {Pesce}, \citenamefont {Petrov}, \citenamefont {Piepmeier},
  \citenamefont {Roberts}, \citenamefont {Robinson}, \citenamefont {Shieler},
  \citenamefont {Small}, \citenamefont {Spellmeyer}, \citenamefont {Tiede},
  \citenamefont {Verniero}, \citenamefont {Wang}, \citenamefont {Wielgus},
  \citenamefont {Wollack}, \citenamefont {Wong},\ and\ \citenamefont
  {Yang}}]{10.1117/12.2630313}%
  \BibitemOpen
  \bibfield  {author} {\bibinfo {author} {\bibfnamefont {P.}~\bibnamefont
  {Kurczynski}}, \bibinfo {author} {\bibfnamefont {M.~D.}\ \bibnamefont
  {Johnson}}, \bibinfo {author} {\bibfnamefont {S.~S.}\ \bibnamefont
  {Doeleman}}, \bibinfo {author} {\bibfnamefont {K.}~\bibnamefont {Haworth}},
  \bibinfo {author} {\bibfnamefont {E.}~\bibnamefont {Peretz}}, \bibinfo
  {author} {\bibfnamefont {T.~K.}\ \bibnamefont {Sridharan}}, \bibinfo {author}
  {\bibfnamefont {B.}~\bibnamefont {Bilyeu}}, \bibinfo {author} {\bibfnamefont
  {L.}~\bibnamefont {Blackburn}}, \bibinfo {author} {\bibfnamefont
  {D.}~\bibnamefont {Boroson}}, \bibinfo {author} {\bibfnamefont
  {A.}~\bibnamefont {Brosius}}, \bibinfo {author} {\bibfnamefont
  {R.}~\bibnamefont {Butler}}, \bibinfo {author} {\bibfnamefont
  {D.}~\bibnamefont {Caplan}}, \bibinfo {author} {\bibfnamefont
  {K.}~\bibnamefont {Chatterjee}}, \bibinfo {author} {\bibfnamefont
  {P.}~\bibnamefont {Cheimets}}, \bibinfo {author} {\bibfnamefont
  {D.}~\bibnamefont {D'Orazio}}, \bibinfo {author} {\bibfnamefont
  {T.}~\bibnamefont {Essinger-Hileman}}, \bibinfo {author} {\bibfnamefont
  {P.}~\bibnamefont {Galison}}, \bibinfo {author} {\bibfnamefont
  {R.}~\bibnamefont {Gamble}}, \bibinfo {author} {\bibfnamefont
  {S.}~\bibnamefont {Hadar}}, \bibinfo {author} {\bibfnamefont
  {T.}~\bibnamefont {Hoerbelt}}, \bibinfo {author} {\bibfnamefont
  {H.}~\bibnamefont {Jiao}}, \bibinfo {author} {\bibfnamefont {J.}~\bibnamefont
  {Kauffmann}}, \bibinfo {author} {\bibfnamefont {R.}~\bibnamefont {Lafon}},
  \bibinfo {author} {\bibfnamefont {C.-P.}\ \bibnamefont {Ma}}, \bibinfo
  {author} {\bibfnamefont {G.}~\bibnamefont {Melnick}}, \bibinfo {author}
  {\bibfnamefont {N.~R.}\ \bibnamefont {Newbury}}, \bibinfo {author}
  {\bibfnamefont {S.}~\bibnamefont {Noble}}, \bibinfo {author} {\bibfnamefont
  {D.}~\bibnamefont {Palumbo}}, \bibinfo {author} {\bibfnamefont
  {L.}~\bibnamefont {Paritsky}}, \bibinfo {author} {\bibfnamefont
  {D.}~\bibnamefont {Pesce}}, \bibinfo {author} {\bibfnamefont
  {L.}~\bibnamefont {Petrov}}, \bibinfo {author} {\bibfnamefont
  {J.}~\bibnamefont {Piepmeier}}, \bibinfo {author} {\bibfnamefont {C.~J.}\
  \bibnamefont {Roberts}}, \bibinfo {author} {\bibfnamefont {B.}~\bibnamefont
  {Robinson}}, \bibinfo {author} {\bibfnamefont {C.}~\bibnamefont {Shieler}},
  \bibinfo {author} {\bibfnamefont {J.}~\bibnamefont {Small}}, \bibinfo
  {author} {\bibfnamefont {N.}~\bibnamefont {Spellmeyer}}, \bibinfo {author}
  {\bibfnamefont {P.}~\bibnamefont {Tiede}}, \bibinfo {author} {\bibfnamefont
  {J.}~\bibnamefont {Verniero}}, \bibinfo {author} {\bibfnamefont
  {J.}~\bibnamefont {Wang}}, \bibinfo {author} {\bibfnamefont {M.}~\bibnamefont
  {Wielgus}}, \bibinfo {author} {\bibfnamefont {E.}~\bibnamefont {Wollack}},
  \bibinfo {author} {\bibfnamefont {G.~N.}\ \bibnamefont {Wong}}, \ and\
  \bibinfo {author} {\bibfnamefont {G.}~\bibnamefont {Yang}},\ }in\ \href
  {\doibase 10.1117/12.2630313} {\emph {\bibinfo {booktitle} {Space Telescopes
  and Instrumentation 2022: Optical, Infrared, and Millimeter Wave}}},\ Vol.\
  \bibinfo {volume} {12180},\ \bibinfo {editor} {edited by\ \bibinfo {editor}
  {\bibfnamefont {L.~E.}\ \bibnamefont {Coyle}}, \bibinfo {editor}
  {\bibfnamefont {S.}~\bibnamefont {Matsuura}}, \ and\ \bibinfo {editor}
  {\bibfnamefont {M.~D.}\ \bibnamefont {Perrin}}},\ \bibinfo {organization}
  {International Society for Optics and Photonics}\ (\bibinfo  {publisher}
  {SPIE},\ \bibinfo {year} {2022})\ p.\ \bibinfo {pages} {121800M}\BibitemShut
  {NoStop}%
\bibitem [{\citenamefont {Moore}\ \emph {et~al.}(2015)\citenamefont {Moore},
  \citenamefont {Cole},\ and\ \citenamefont {Berry}}]{Moore:2014lga}%
  \BibitemOpen
  \bibfield  {author} {\bibinfo {author} {\bibfnamefont {C.~J.}\ \bibnamefont
  {Moore}}, \bibinfo {author} {\bibfnamefont {R.~H.}\ \bibnamefont {Cole}}, \
  and\ \bibinfo {author} {\bibfnamefont {C.~P.~L.}\ \bibnamefont {Berry}},\
  }\href {\doibase 10.1088/0264-9381/32/1/015014} {\bibfield  {journal}
  {\bibinfo  {journal} {Class. Quant. Grav.}\ }\textbf {\bibinfo {volume}
  {32}},\ \bibinfo {pages} {015014} (\bibinfo {year} {2015})},\ \Eprint
  {http://arxiv.org/abs/1408.0740} {arXiv:1408.0740 [gr-qc]} \BibitemShut
  {NoStop}%
\end{thebibliography}%
\end{document}